\documentclass[12pt]{iopart}

\usepackage{cite,epsfig,graphicx,amssymb}
\begin{document}

\setcounter{page}{1}
\title[Predictions for the LHC heavy ion programme]{Predictions for the LHC heavy ion programme}

\author{Nicolas Borghini\dag\ and Urs Achim Wiedemann\ddag}

\address{\dag\ Fakult\"at f\"ur Physik, Universit\"at Bielefeld, 
  Postfach 100131, D-33501 Bielefeld, Germany}
\address{\ddag\ CERN, Department of Physics, Theory Division, CH-1211 Geneva 23,
Switzerland}
\ead{borghini@physik.uni-bielefeld.de, Urs.Wiedemann@cern.ch}
\begin{abstract}
Apparently universal trends have been observed in relativistic 
nucleus-nucleus collisions up to RHIC energies. Here, we review these trends and we 
discuss their agnostic extrapolation to heavy ion collisions at the LHC.  
\end{abstract}

\pacs{12.38.Mh, 25.75.Nq}

\begin{center}
{\sl Invited topical review for \jpg}\\
\end{center}

\maketitle

\section{Introduction}
\label{sec:intro}

The main goal of ultra-relativistic heavy ion physics is to test the properties of matter, produced in
nucleus-nucleus collisions at the highest energy densities accessible in the laboratory, and to 
develop an understanding of these properties from first principles of the fundamental theory of 
strong interactions,  Quantum Chromodynamics (QCD). To this end, particle production in heavy
ion collisions and its dependence on the formation of a dense system can be studied experimentally
as a function of a large number of variables. These variables include the kinematic ones (such as 
centre-of-mass energy, transverse momentum and rapidity), as well as variables specific to heavy 
ion collisions, which typically control the size and shape of the collision region (such as the impact 
parameter  or the nuclear number $A$ of the colliding nuclei). 

Over the last two decades, relativistic heavy ion collisions have been studied experimentally
at increasingly higher centre-of-mass energies at the Brookhaven Alternating Gradient
Synchrotron AGS ($\sqrt{s_{_{NN}}} < 5$ GeV), the CERN Super Proton Synchrotron SPS
($\sqrt{s_{_{NN}}} \leq 20$ GeV) and the Brookhaven Relativistic Heavy Ion Collider RHIC
($\sqrt{s_{_{NN}}} \leq 200$ GeV). As discussed in this article, the data collected in these 
experimental campaigns display remarkable generic trends as a function of system size and 
kinematic variables.  

The Large Hadron Collider LHC at CERN will study heavy ion collisions at a centre-of-mass
energy $\sqrt{s_{_{NN}}} = 5.5$ TeV, which is almost a factor 30 higher than the maximal
collision energy at RHIC. There has been a lot of work in recent years on benchmarking
model calculations of heavy ion collisions to RHIC data and extrapolating them to the higher
LHC energies\footnote{A comprehensive update on these efforts will be given in the proceedings of
a recent CERN Theory Institute workshop~\cite{CERNTHinst}.}. The present topical review does not aim 
at such a comprehensive summary. Rather, the aim of this review is to identify the 
generic trends in the existing data and to discuss the consequences for our understanding
of heavy ion collisions if these trends should persist or should fail to persist at the LHC. For those classes
of measurements, where guidance from existing data is scarce (for instance for measurements
at high-$p_T$ or forward rapidity, where LHC is unique), we shall focus on generic features in the 
current model calculations, and discuss how they are expected to manifest themselves at the LHC.

We believe that the identification of generic trends in the data and their agnostic extrapolation to 
LHC energies may help to sharpen our view on what is expected at LHC energies and what
constitutes a surprise. In addition, many of the generic trends listed in this article did not
yet find a fully satisfactory explanation. If they persist at the LHC, this would indicate that they
should not be discarded as mere numerical coincidences, but should find an explanation in a
future, more complete theory.

\section{Multiplicity distributions}
\label{s:multiplicity}

In $e^+e^- \to q\, \bar{q} \to X$, where the partons produced initially carry perturbatively high 
virtuality, main characteristics of the longitudinal and transverse multiplicity distributions can 
be understood quantitatively from the dynamics of the perturbative parton 
shower (see e.g.~\cite{Dokshitzer:1992jv}), despite
uncertainties in the modelling of hadronization. In contrast, 
in hadronic collisions, event multiplicities and multiplicity distributions are dominated by 
processes involving non-perturbatively small momentum transfers; there are many models 
but an understanding of multiplicity distributions based on first principles is missing. Even in 
proton-proton collisions, the extrapolation
of the charged particle multiplicity per unit rapidity ${\rm d}N_{\rm ch}/{\rm d}y$ from the Tevatron 
($\sqrt{s_{_{NN}}} = 1.8$ TeV) to the LHC ($\sqrt{s_{_{NN}}} = 14$ TeV) leads to results
which vary by a factor 2 for models successful up to Tevatron energies~\cite{Butterworth:2004is}.  
For nucleus-nucleus collisions, the uncertainties in the predictions of minimum bias event 
multiplicities are of comparable magnitude. For instance, prior to the start-up of RHIC, model extrapolations of ${\rm d}N_{\rm ch}/ {\rm d}y$ from the SPS centre-of-mass energy for Pb-Pb collisions
($\sqrt{s_{_{NN}}} = 17$ GeV) to Au-Au collisions at RHIC ($\sqrt{s_{_{NN}}} = 200$ GeV)
varied by a factor 2 approximately~\cite{Armesto:2000xh,Eskola:2001vs}. RHIC data lie
at the lower end of the predicted range. 
This in turn has narrowed the range of predictions 
for the LHC.

\begin{figure}[t]\epsfxsize=10.cm
\centerline{\epsfbox{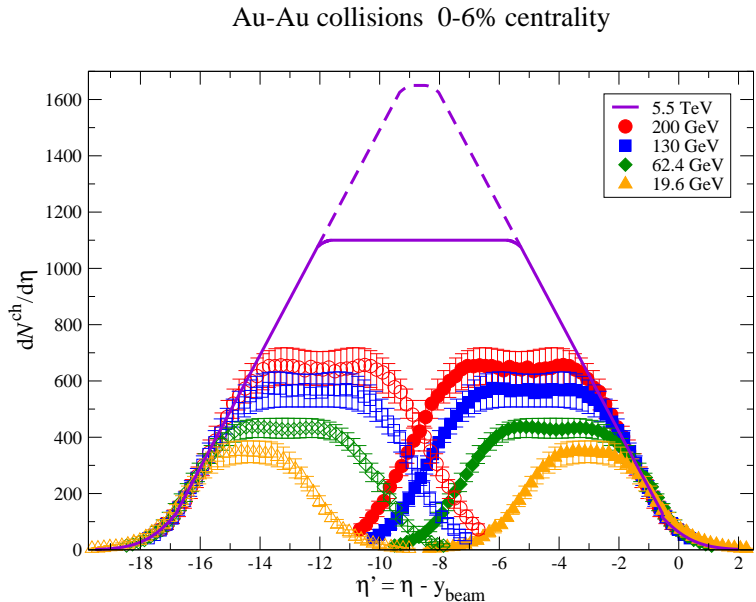}}
\caption{Pseudorapidity distribution of charged particle production in Au-Au collisions at
different centre-of-mass energies. Data are plotted in the rest frame of one of the colliding nuclei
(full symbols), and mirrored at LHC mid-rapidity (open symbols). Agnostic extrapolations to the
LHC are based on assuming limiting fragmentation and i) the saturation ansatz (\ref{eq1})  (dashed line), 
or ii) a self-similar trapezoidal shape of the multiplicity distribution (solid line). Data from~\cite{Back:2002wb,Back:2005hs}.
}\label{fig1}
\end{figure}

The lack of a fundamental understanding of multiparticle production in hadronic collisions
is in marked contrast to several characteristic features, which persist over many orders of 
magnitude in $\sqrt{s_{_{NN}}}$~\cite{Busza:2004mc}:
\begin{enumerate}
	\item \underline{Extended longitudinal scaling (limiting fragmentation~\cite{Benecke:1969sh}).}\\
	Pseudorapidity distributions\footnote{Pseudorapidity $\eta \equiv \tanh^{-1}p_l/p$ is for many purposes a
good approximation of rapidity $y \equiv \tanh^{-1}p_l/E$. However, since multiplicity distributions
are dominated by particles at small transverse momentum, there are visible differences:
${\rm d}N^{AA}_{\rm ch}/{\rm d}\eta$ is of trapezoidal shape (see figure~\ref{fig1}), while 
${\rm d}N^{AA}_{\rm ch}/{\rm d}y$ is of Gaussian shape. So, figure~\ref{fig1} does {\it not} imply a 
rapidity plateau. For $A$-$A$ collisions, multiplying ${\rm d}N/{\rm d}\eta$ at $\eta=0$ by the
conversion factor $\approx 1.1$ turns out to be a good estimate for ${\rm d}N/{\rm d}y$ at $y=0$.}, plotted in the rest frame of one of the colliding hadrons,
	fall on a universal, energy-independent limiting curve in the projectile fragmentation region.
	The region within which this limiting curve is valid, increases with energy, see figure~\ref{fig1}. 
	This is in contrast to the expectation that at high energies a boost-invariant plateau would
	develop around mid-rapidity. 
	\item \underline{Factorization of $\sqrt{s_{_{NN}}}$ and centrality/$A$-dependence.}\\
	For all processes at a given centre-of-mass energy, the pseudorapidity distribution is the same
	basic distribution adjusted for the number of participants in the two colliding systems.
	The $\sqrt{s_{_{NN}}}$- and $N_{\rm part}$-dependences of ${\rm d}N^{AA}_{\rm ch}/{\rm d}\eta$
	factorize.
\end{enumerate}
%

The apparent universality of these observations 
motivates an agnostic extrapolation to the LHC:
If $A$-$A$ data at LHC follow the same limiting fragmentation curve and if the trapezoidal shape
of pseudo-rapidity distributions persists, then one expects for ${\rm d}N_{\rm AA}^{\rm ch}/{\rm d}\eta$
at LHC the solid line in figure~\ref{fig1}. This curve implies ${\rm d}N_{\rm PbPb}^{\rm ch}/{\rm d}\eta \sim 1100$. More generally, the persistence of extended longitudinal scaling in the high energy limit
implies that at mid-rapidity, ${\rm d}N/{\rm d}y$ can grow at most logarithmically with $\sqrt{s}$,
except if there is a novel mechanism due to which the curvature of ${\rm d}N/{\rm d}y$ changes its
sign {\it twice} between $y=0$ and the fragmentation region.

Essentially all models of multiplicity distributions predict a power-law increase with  
$\sqrt{s_{_{NN}}}$. This is a rather generic consequence of perturbative particle production 
mechanisms, which become more important with increasing $\sqrt{s}$. 
However, the power-law dependence of naive perturbative implementations is too strong
to be reconciled with RHIC data --- this is arguably the main lesson learnt from the failure
of many models at RHIC. Saturation models have received much attention recently, since
they offer a fundamental reason for the very weak $\sqrt{s}$-dependence of event multiplicities, 
namely the taming of the perturbative rise due to density-dependent non-linear parton evolution.
Still, saturation models assume that multiplicity distributions at ultra-relativistic energies are 
calculable within perturbation theory, since they are governed by a perturbatively high, $\sqrt{s}$- 
and $A$-dependent momentum (saturation) scale $Q_{{\rm sat}, A}^2 \propto \sqrt{s}^\lambda$. 
They predict essentially, that multiplicities at mid-rapidity rise $\propto Q_{{\rm sat}, A}^2$ times 
transverse area. This leads e.g. to the pocket formula~\cite{Armesto:2004ud}
\begin{equation}
\frac{2}{N_{\rm part}}\frac{{\rm d}N^{AA}_{\rm ch}}{{\rm d}\eta} \Bigg\vert_{\eta \sim 0}
	= N_0\, \sqrt{s_{NN}\mbox{ [in GeV]}}^{\,\lambda} N_{\rm part}^{\frac{1-\delta}{3\delta}}\, .
\label{eq1}
\end{equation}
Here $N_0 = 0.47$ is fixed by fitting to RHIC multiplicity distributions. 
The factors $\lambda = 0.288$ and $\delta = 0.8$ are constrained by fitting data
on $eA$ inelastic scattering. 
In this sense, the $\sqrt{s}$- and $A$-dependences of (\ref{eq1}) are predicted.
The pocket formula (\ref{eq1}) provides an explicit realization of the 
factorization property (ii) stated above, and accounts satisfactorily for the 
$\sqrt{s}$- and $A$-dependences of charged multiplicity distributions from SPS 
to maximal RHIC energies~\cite{Armesto:2004ud}. 
For central Pb-Pb collisions at the LHC, equation~(\ref{eq1}) leads to 
${\rm d}N_{\rm PbPb}^{\rm ch}/{\rm d}\eta \sim 1650$, which corresponds to the 
maximum of the dashed line in figure~\ref{fig1}. 
An alternative model implementation of saturation physics ideas arrives at 
${\rm d}N_{\rm PbPb}^{\rm ch}/{\rm d}\eta \sim 2200$~\cite{Kharzeev:2004if}.
A similar value is also predicted by the EKRT final state saturation 
model~\cite{Eskola:1999fc}, which arrives at a $\sqrt{s_{_{NN}}}^{0.38}$-dependence 
of the charged multiplicity at mid-rapidity.
To arrive at a $\ln \sqrt{s_{_{NN}}}$-dependence in such schemes, one would have to 
invoke a mechanism, which amputates brutally the power-law tail in the spectrum 
for $p_T^2 > Q_s^2$ (see e.g. the discussion of D. Kharzeev in reference~\cite{CERNTHinst}).

A complete list of model predictions is beyond the scope of this article. 
We emphasize, however, that none of these predictions can be reconciled 
with the assumption that the so-far universal extended longitudinal scaling 
persists at LHC energies. 
As seen from figure~\ref{fig1}, a distribution matching 
${\rm d}N_{\rm PbPb}^{\rm ch}/{\rm d}\eta \sim 2200$ at mid-rapidity has to fall off
significantly steeper than what is consistent with limiting fragmentation. 
Also, the prediction ${\rm d}N_{\rm PbPb}^{\rm ch}/{\rm d}\eta \sim 1650$, while 
being close to the maximum of what is consistent with limiting fragmentation, 
appears to deviate characteristically from the trapezoidal shape of all 
pseudo-rapidity distributions of charged multiplicity measured so far. 
 
To summarize and generalize this discussion: 
Either, the apparently universal ``structure'' seen in multiparticle production 
data at lower energies~\cite{Busza:2004mc} is violated at the LHC. 
Then, this violation is likely to provide highly discriminatory constraints on 
the dynamics underlying multiparticle production.
Or, the naive extrapolations of this structure to the LHC are confirmed. 
Then, the central dynamical ideas advocated as explanations for the tamed growth
of multiplicities up to RHIC energies will have to be revisited. 
So, starting with the first day of operation, data from the LHC are likely to 
have profound consequences for our understanding of the matter produced in 
nucleus-nucleus collisions at the LHC {\it and\/} at RHIC.

\section{Hadrochemistry}
\label{s:hadrochemistry}

\begin{figure}[t]
\centerline{\includegraphics*[width=11cm]{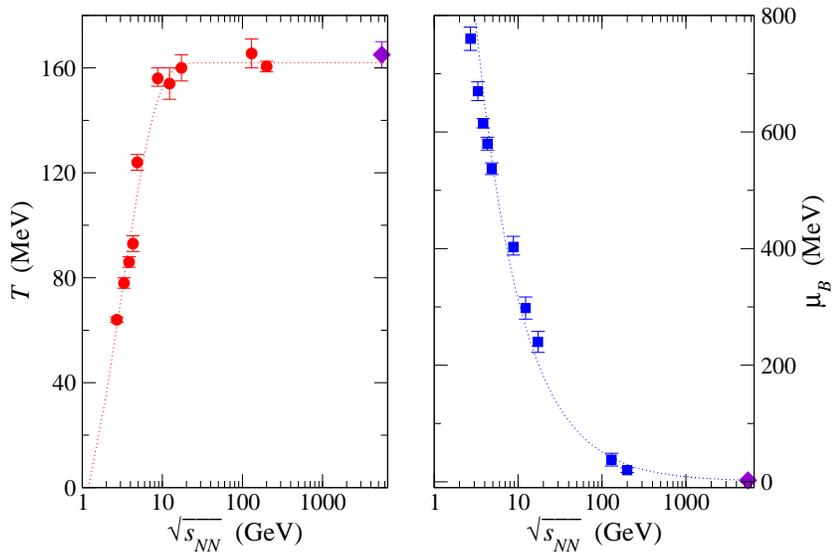}}
\caption{Thermal model fits at mid-rapidity of the hadrochemical freeze-out 
  temperature $T$ and the baryochemical potential $\mu_B$ as a function of the 
  centre-of-mass energy $\sqrt{s_{_{NN}}}$.
  Data points up to RHIC energies are taken from reference~\cite{Andronic:2005yp}. 
  Data points at $\sqrt{s_{_{NN}}}=5.5$~TeV are based on simple extrapolations of 
  the observed tendency.}
\label{fig2}
\end{figure}

The relative abundance of identified hadron species in heavy ion collisions 
follows a statistical (``thermal'') distribution pattern over a very broad energy 
range from SIS/GSI ($\sqrt{s_{_{NN}}} \approx 2$~GeV) up to RHIC 
($\sqrt{s_{_{NN}}} = 200$~GeV)~\cite{Braun-Munzinger:2003zd}. 
For particle species, for which global constraints (such as total charge or 
flavour conservation) are statistically unimportant because of sufficiently large
event multiplicities, particle ratios are well described by the grand canonical 
ensemble of a hadron resonance gas. 
The only parameters of this model are the temperature $T$ at chemical 
decoupling, and the baryon chemical potential $\mu_B$.
Fits to particle ratios reveal a characteristic $\sqrt{s_{_{NN}}}$-dependence of 
these parameters, see figure~\ref{fig2}. 
The baryon chemical potential decreases by almost an order of magnitude from 
$\mu_B\sim 250$~MeV at the SPS to $\mu_B\sim 20\,$-\,40~MeV at RHIC and is generally
assumed to be very small ($\mu_B\ll 10$~MeV) at LHC mid-rapidity. 
This reflects the expectation that due to the large difference between 
projectile and mid-rapidity at the LHC, the mid-rapidity region is almost 
net-baryon free. 
The chemical decoupling temperature approaches a $\sqrt{s_{_{NN}}}$-independent 
limiting temperature $T\simeq 160-170$~MeV, which seems to be almost reached at 
RHIC and which is expected to persist at the LHC. 
These agnostic extrapolations to LHC energies have been used to predict the 
ratios of more than twenty hadron species, and there are computer codes 
implementing these model assumptions~\cite{Cleymans:2006xj,Wheaton:2004qb,%
  Wheaton:2004vg,Torrieri:2006xi}.
Early works on hadronic abundances used a strangeness saturation 
parameter~\cite{Letessier:2002gp} to account for enhanced strangeness production
with increasing $\sqrt{s_{_{NN}}}$. 
This followed the idea that with increasing $\sqrt{s_{_{NN}}}$ a gluon-rich initial 
system is formed, which is more efficient in producing strangeness. 
In contrast, the above-mentioned models do not fix the strangeness content with 
an additional parameter. 
These models explain strangeness increase by a {\it suppression\/} of strange 
particles in low-multiplicity systems created at lower $\sqrt{s_{_{NN}}}$. 
This suppression is due to exact conservation laws, whose implementation leads 
to deviations from the grand-canonical limit~\cite{Braun-Munzinger:2003zd}. 
The resulting canonical suppression factors are known analytically. 

At RHIC and at the LHC, canonical suppression is unimportant for strangeness, 
which is produced abundantly. 
On the other hand, canonical suppression is important for open charm~\cite{%
  Andronic:2003zv} and bottom production at the LHC. 
In this general sense, charm and bottom are likely to play at collider energies 
a similar role as strangeness played during the fixed-target era of heavy-ion 
physics. 
There is, however, one important difference: because of their large mass and 
small production cross section at thermal energies, the thermal production of 
charm and bottom quarks is disfavoured. 
Thus, except for the very first proposals of thermal charm production~\cite{%
  Gazdzicki:1999rk}, models assume normally a production of heavy-flavoured 
quarks at perturbative rates, followed by a hadrochemical distribution of 
heavy-flavoured resonances according to the statistical model of 
hadroproduction~\cite{Andronic:2003zv}. 
In this approach, the centrality dependence of charm-flavoured particle ratios 
tests the transition from a canonical to a grand-canonical description. 

The models discussed so far implement the idea of statistical hadronization, but
they do {\it not\/} specify the dynamics leading to the hadronic final state. 
In particular, although the hadrochemical freeze-out temperature and 
baryochemical potential shown in figure~\ref{fig2} match within errors the QCD 
phase-space boundary determined in lattice QCD, the question whether this 
agreement implies the existence of dynamical thermalization processes lacks a 
more detailed support~\cite{Braun-Munzinger:2003zz}. 
We conclude this section by listing three possibilities of how heavy-ion 
collisions at the LHC may help to elucidate the microscopic dynamics underlying 
hadronic abundances:
First, it has been conjectured that charmonium production at the LHC is strongly
enhanced above the perturbatively expected rates due to the recombination of 
$c$ and $\bar{c}$ quarks originating in distinct hard partonic 
interactions~\cite{Thews:2000rj}. 
Establishing such a novel charmonium production mechanism could provide a strong
indication that thermalization processes affect hadrochemical distributions. 
This is so, since transport of exogamous $c$- and $\bar{c}$-partners is likely 
to be necessary to make their coalescence possible. 
Second, we mention in passing a proposal that the abundances of strange hadrons 
could exceed significantly grand-canonical predictions. 
This is seen in models in which a {\it sudden} hadronization of an equilibrated 
plasma leads to strangeness over-saturation in the hadronic phase~\cite{%
  Rafelski:2005jc}. 
While arguably speculative, this example illustrates that the extrapolation of 
(standard) statistical model predictions to the LHC can serve as a powerful 
baseline on top of which novel dynamical effects may be established.
Third, the masses and widths of the 
resonances entering the statistical operator of a resonance gas model may 
receive significant medium-modifications. This may lead to deviations of some broad 
resonances (such as $\rho$-mesons) from the grand canonical ensemble.
While not specific for the LHC, it is likely that any confirmation and 
extended systematics of such deviations will provide constraints on the 
in-medium dynamics of hadronic resonances near freeze-out.

\section{Transverse-momentum spectra at low $p_T$}
\label{s:pTspectra}

The transverse momentum dependence of identified single inclusive hadron
spectra has been studied in heavy ion collisions from SIS/GSI up to 
RHIC energies. Here, we follow the common practice of discussing these
spectra as a function of their transverse mass $m_T=\sqrt{m^2 + p_T^2}$,
rather than their transverse momentum. 
The $m_T$-distributions are commonly characterized by a mono-exponential fit
${\rm d}N/{\rm d}m_T \propto m_T^{3/2} \exp\left[-m_T/T_{\rm inv}\right]$~\cite{Lee:1990sk}, 
where the inverse-slope parameter 
$T_{\rm inv}$ characterizes the steepness of the distributions. The fit value
$T_{\rm inv}$ can depend on the $m_T$-range over which the fit is performed, 
but despite the ensuing uncertainties, the following generic features can be stated:
\begin{enumerate}
\item \underline{Linear mass-dependence of $T_{\rm inv}$.}\\
  At given $\sqrt{s_{_{NN}}}$, the inverse-slope parameters of pions, 
  kaons, and protons increase {\em approximately linearly\/} with the particle rest mass.
  In particular $T_{{\rm inv,}\,\pi^\pm}<T_{{\rm inv,}\,K^\pm}<T_{{\rm inv,}\,p}$. 
  The hadrons, which do not follow this systematics, such a the multistrange $\Xi$, the
   $\Omega$ or the $J/\Psi$, show roughly the same value $T_{\rm inv}$. 
   A seemingly common denominator of these latter hadrons is their relatively small
  interaction cross sections with the expected constituents of the 
  medium~\cite{Xu:2001zj,Antinori:2004va}.
\item \underline{Increase of $T_{\rm inv}$ with $\sqrt{s_{_{NN}}}$.}\\
  The spectra become flatter with increasing $\sqrt{s_{_{NN}}}$, so the 
  inverse slope parameter $T_{\rm inv}$ increases. Also, the mass-dependence
  of $T_{\rm inv}$ for pions, kaons and protons increases with $\sqrt{s_{_{NN}}}$~\cite{%
    Xu:2001zj}.
\end{enumerate}
The inverse-slope parameter $T_{\rm inv}$ has been interpreted as
a blue-shifted temperature, resulting from a combination of thermal emission 
from a source of temperature $T_{\rm f.o.}$ at freeze-out, and the collective 
motion with average transverse velocity $\langle\beta_T\rangle$ of this
particle emitting source. While $T_{\rm f.o.}$ and $\langle\beta_T\rangle$ are
difficult to disentangle on the basis of single inclusive spectra alone, 
two-particle correlations may be used to separate both contributions~\cite{Wiedemann:1999qn}. 
The increase of $T_{\rm inv}$ with particle rest mass [point (i)]
is consistent with the interpretation of  $T_{\rm inv}$ as a blue-shifted temperature~\cite{Lee:1990sk,Retiere:2003kf}, if the particles
decouple at the time of freeze-out. On the other hand, particles which decouple 
earlier due to their smaller cross sections would acquire less radial flow, and thus 
have a smaller $T_{\rm inv}$. 
In this radial-flow picture, the growing difference between inverse-slope 
parameters as $\sqrt{s_{_{NN}}}$ increases [point (ii)] is ascribed to the growth
in the average transverse velocity $\langle\beta_T\rangle$.

The above features emerge naturally in fluid-dynamics models 
without~\cite{Teaney:2001av,Kolb:2003dz} or with~\cite{Baier:2006um} viscous
corrections. We are not aware of full fluid-dynamic studies of these spectra at
LHC energy. However, there are extrapolations of the fit parameters
$T_{\rm f.o.}$ and $\langle\beta_T\rangle$, extracted from the approximately 
exponential $m_T$-spectra~\cite{Xu:2001zj}. In these studies, the kinetic freeze-out 
temperature saturates at a value $T_{\rm f.o.}\approx 120$~MeV at SPS energy, 
which may be expected to persist up to the LHC. 
The parameter $\langle\beta_T\rangle$ is found to increase with $\sqrt{s_{_{NN}}}$; 
it reaches  $\langle\beta_T\rangle\approx 0.55$ at RHIC, but
it remains unclear whether this increase is smooth~\cite{Xu:2001zj}. 

The statements made here are obtained within a blast-wave model~\cite{Xu:2001zj,Retiere:2003kf} 
or within a fluid-dynamic picture supporting such a model. They generally indicate an increase
of $T_{\rm inv}$ with $\sqrt{s_{_{NN}}}$. Here, the real physical
issue is to assess whether it is really a collective hydrodynamic mechanism, 
which determines the energy-dependence of $T_{\rm inv}$. Since the slope of spectra
evolves significantly with $\sqrt{s_{_{NN}}}$ also for more elementary proton-proton 
collisions, in which flow is either absent or at least different, it is not straightforward to 
disentangle flow effects from other mechanisms. This makes quantitative comparisons
difficult and leads to a lack of precision in extrapolations to the LHC. We now turn to 
the azimuthal dependence of single inclusive transverse momentum spectra, where 
these difficulties appear to be less severe.

\section{Azimuthal anisotropy in low-$p_T$ particle production}
\label{s:anisotropic_flow}

In non-central collisions, the impact parameter selects a preferred direction 
in the transverse plane, breaking the rotational symmetry around the beam axis.
Consequently, observables can depend on azimuth, measured with respect to the 
reaction plane spanned by the impact parameter and the beam direction. 
The azimuthally-dependent effect most widely studied both experimentally, over 
the whole range of available collision energies, and theoretically, is the 
azimuthal anisotropy in particle production, often referred to as ``anisotropic 
(transverse) flow''~\cite{Kolb:2003dz}. 
A non-vanishing anisotropic flow exists only if the particles measured in the 
final state depend not only on the physical conditions realized locally at their
production point, but if particle production does also depend on the global 
event geometry. 
In a relativistic local theory, this non-local information can only emerge as a 
collective effect, requiring interactions between many degrees of freedom, 
localized at different points in the collision region. 
In this general sense, anisotropic flow is a particularly unambiguous and strong
manifestation of collective dynamics in heavy-ion collisions. 
Here, we discuss its features for hadrons at low-$p_T$. 

Azimuthal anisotropies in particle production are most conveniently 
characterized by performing a Fourier expansion of the single-particle 
distribution
\begin{equation}
\fl
\frac{{\rm d}N}{{\rm d}^2{\bf p}_t\,{\rm d} y} = 
\frac{1}{2\pi}\frac{{\rm d}N}{p_T\,{\rm d}p_T\,{\rm d}y}
\left[ 1 + 2v_1\cos(\phi-\Phi_R) + 2v_2\cos 2(\phi-\Phi_R) + \cdots \right],
\label{eq2}
\end{equation}
where $\Phi_R$ denotes the azimuth (in the laboratory frame) of the reaction 
plane.
The coefficients $v_n=\langle \cos n(\phi-\Phi_R) \rangle$, where angular 
brackets denote an average over many particles and events, quantify the 
asymmetry. 
These coefficients are studied at all available energies as a function of 
transverse momentum, (pseudo)rapidity, centrality of the collision and for 
various identified particle species. 
\begin{figure}[t]
\centerline{\includegraphics*[width=10cm]{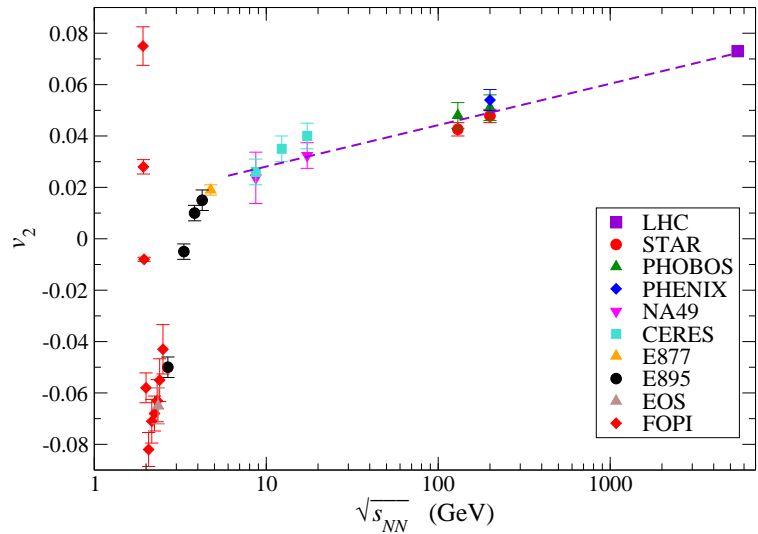}}
\caption{$\sqrt{s_{_{NN}}}$-excitation function of $v_2(y\!=\!0)$ in mid-central 
  collisions. 
  Data are taken from the compilation in reference~\cite{Alessandro:2006yt}.}
\label{fig:flow2}
\end{figure}

``Elliptic flow'', the second Fourier coefficient $v_2$ is the best studied one. 
A positive (resp. negative) value of $v_2$ indicates an excess of particle 
production in (resp. orthogonal to) the reaction plane. 
The dependence of $v_2$ on centre-of-mass energy is known over three orders of 
magnitude, see figure~\ref{fig:flow2}. 
It can be understood qualitatively in terms of the following simple picture of a
collective dynamics: 
At $\sqrt{s_{_{NN}}} < 2$~GeV, the incoming nuclei transfer angular momentum to 
the nuclear matter in the overlap zone. 
The fast-rotating ``compound nucleus'' thus formed emits fragments, which lie 
preferentially in the reaction plane ($v_2>0$). 
As the centre-of-mass energy increases, the nuclear matter in the almond-shaped 
overlap region of the incoming nuclei is increasingly compressed in the 
collision. 
However, the parts of the nuclei that lie outside this overlap region
(``spectators'') block the way for the compressed matter to expand within the 
reaction plane. 
They {\it squeeze-out\/} the compressed matter orthogonal to the reaction plane 
($v_2<0$). 
Further increasing $\sqrt{s_{_{NN}}}$, the spectators are then fast enough to free 
the way, leaving behind at mid-rapidity an almond-shaped azimuthally asymmetric 
region of dense QCD matter. 
This spatial asymmetry implies unequal pressure gradients in the transverse 
plane, with a larger gradient in the reaction plane (``in-plane'') than 
perpendicular to it. 
As a consequence of the subsequent multiple interaction between many degrees of 
freedom, this spatial asymmetry leads to an anisotropy in momentum space: 
the final particle transverse momenta are more likely to be in-plane than 
``out-of-plane'', hence $v_2>0$, as predicted in~\cite{Ollitrault:1992bk}. 

\begin{figure}[t]
\centerline{\includegraphics*[width=10cm]{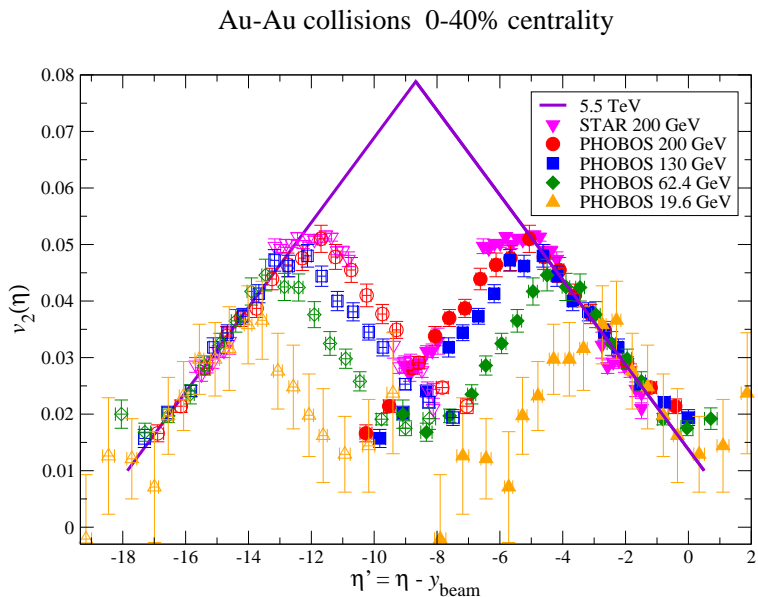}}
\caption{The elliptic flow $v_2$, averaged over centrality (0\%-40\%), at 
  various collision energies. Data (full symbols) from PHOBOS~\cite{Back:2004zg}
  and STAR~\cite{Adams:2004bi} are plotted as a function of $\eta-y_{\rm beam}$ 
  and reflected (open symbols) across the LHC $-y_{\rm beam}$ value.}
\label{fig:flow1}
\end{figure}

The momentum space asymmetries measured at collider energies are relatively 
large. 
Since the prefactor of the cosine term in equation~(\ref{eq2}) is $2v_2$, a 
$p_T$-averaged value $v_2=0.05$ corresponds to a 20\% variation of the average 
particle yield as a function of the angle with respect to the reaction plane.
At high $p_T$, where second harmonics at RHIC approached values as large as
$v_2=0.2$, there are more than twice the number of particles emitted in the 
reaction plane than out-of-plane. 
Elliptic flow is an abundant and very strong manifestation of collectivity, 
which shows remarkable generic trends:

\begin{enumerate}
\item 
The $p_T$-integrated $v_2(\eta)$ shows 
\underline{extended longitudinal scaling}~\cite{Back:2004zg}. \\
In contrast to ${\rm d}N/{\rm d}\eta$, $v_2(\eta)$ is not trapezoidal 
but triangular, see figure~\ref{fig:flow1}\footnote{%
  The $p_T$-averaged value of $v_2$ is dominated by values of the transverse
  momentum close to $\langle p_T\rangle$, so that $v_2(\eta)$ and $v_2(y)$ are 
  similar, in contrast to ${\rm d}N/{\rm d}\eta$ and ${\rm d}N/{\rm d}y$.}.
  As seen clearly from figure~\ref{fig:flow1}, longitudinal scaling of $p_T$-integrated
  $v_2$ persists up to mid-rapidity. 
\item
The \underline{$p_T$-shape of the charged-hadron $v_2$} has a characteristic 
breaking point.\\
At transverse momenta below $p_T\simeq 2$~GeV/$c$, where data are known from SPS and
RHIC, $v_2$ is found to have an approximately linear rise with $p_T$. 
Around $p_T \simeq 2$~GeV/$c$, this rise levels off rather abruptly. 
The energy-dependence of this $p_T$-shape is not fully clarified: 
At low $p_T$, the slope of $v_2(p_T)$ was reported to rise either 
slightly~\cite{Snellings:2003mh} or significantly~\cite{Adler:2004cj} across 
SPS and RHIC energies. 
Also, it was reported~\cite{Adler:2004cj} that the slope of $v_2(p_T)$ saturates
at RHIC energies and is essentially constant between $\sqrt{s_{_{NN}}}=62.4$~GeV 
and 200~GeV. 
In this case, the increase of the $p_T$-averaged $v_2$ would be entirely due to 
the increase in the mean transverse momentum of particles. 
\item  
The $p_T$-dependent $v_2$ of identified hadrons shows \underline{mass-ordering}
at small $p_T$ and displays a \underline{constituent-quark counting rule} at 
intermediate $p_T$.
\\
For a fixed, sufficiently low transverse momentum, SPS and RHIC data show 
generically that $v_2(p_T)$ decreases with increasing mass of the particle 
species. 
Above a critical $p_T\sim 1.5$~GeV/$c$, mass-ordering ceases to be valid and 
$v_2(p_T)$ follows to a good approximation a simple quark counting rule, namely 
that $v_2(p_T/n_q)/n_q$ is roughly independent of the particle 
species~\cite{Voloshin:2002wa}.
\end{enumerate}
%

What are the implications if these trends persist or do not persist at the LHC?\\
First, if longitudinal scaling of $v_2$ persists, then $v_2(\eta)$ grows 
proportional to $\ln\sqrt{s_{_{NN}}}$. 
In this case, one expects $v_2(\eta\!=\!0)\simeq 0.075$ for Pb-Pb collisions in 
mid-central collisions. 
This follows from the extrapolations, shown in figures~\ref{fig:flow2} and 
\ref{fig:flow1}. 
To the best of our knowledge, neither the triangular shape of the rapidity 
dependence of $v_2$, nor the approximately linear $\ln \sqrt{s_{_{NN}}}$-dependence
emerges as a natural consequence of existing dynamical models. 
In particular, extrapolating models of ideal hydrodynamics from RHIC to the LHC,
one arrives at values not exceeding $v_2(\eta\!=\!0)\simeq 0.06$ for event 
multiplicities shown in figure~\ref{fig1}~\cite{Teaney:2001av}.
Also, the proportionality $v_2(y) \propto {\rm d}N/{\rm d}y$ does not 
hold in models presupposing local equilibrium (i.e. the formation of an almost 
perfect fluid). If the matter produced at RHIC mid-rapidity is in local equilibrium, then 
one expects at the LHC deviations from the triangular shape of $v_2(\eta)$ in an
extended region $-3 \lesssim y \lesssim 3$, around mid-rapidity\footnote{%
  However, in models which presuppose incomplete local equilibration~\cite{%
  Heinz:2004et,Bhalerao:2005mm}, one finds $v_2(y) \propto {\rm d}N/{\rm d}y$, 
  thus accounting for the triangular shape of $v_2(\eta)$.}. 
In summary: extrapolations to the LHC of the main dynamical mechanism
advocated to underlie elliptic flow at RHIC (namely perfect fluid dynamics) are at odds
with extrapolations to the LHC of the generic trends observed in elliptic flow measurements
up to RHIC energy. As a consequence, establishing whether these trends persist at the LHC 
provides a novel independent test for our understanding of the properties of matter at the LHC 
{\it and\/} at RHIC. 

Second, comparing measurements of $v_2(p_T)$ from LHC and RHIC at low $p_T$ will
finally allow us to establish to what extent the $p_T$-slope of $v_2(p_T)$ 
changes with $\sqrt{s_{_{NN}}}$. 
This is of interest, since a $\sqrt{s_{_{NN}}}$-independent slope of $v_2(p_T)$ 
would imply e.g. that the increase of the $p_T$-integrated $v_2(\eta)$ with 
$\sqrt{s_{_{NN}}}$ arises solely from the increase of the average transverse 
momentum.
Existing dynamical models of $v_2$ do not invoke the increase of the rms 
transverse momentum $\sqrt{\langle p_T^2}\rangle$ with $\sqrt{s_{_{NN}}}$ observed
in hadronic collisions; 
so, establishing a major role of $\sqrt{\langle p_T^2}\rangle$ in $v_2$ may 
prompt significant revision in our interpretation of elliptic flow. 
Moreover, LHC data for $v_2(p_T)$ at intermediate $p_T$ will test to what extent
the breaking point of $v_2(p_T)$ depends on centre-of-mass energy, collision 
centrality, or choice of nuclei, while existing RHIC data hint at a very small 
sensitivity on these~\cite{Adare:2006ti}. 
In the current discussion of RHIC data, one emphasizes that the approximately 
linear increase of $v_2(p_T)$ at low $p_T$ is consistent with the transverse 
expansion of an almost perfect liquid~\cite{Kolb:2003dz}, 
while the breaking point in $v_2(p_T)$ arises from the rather abrupt onset of 
dissipative effects at higher $p_T$~\cite{Teaney:2003pb}.
However, transport models and dissipative hydrodynamics~\cite{Teaney:2003pb} can
account for the linear increase of $v_2(p_T)$ at small $p_T$ as well, if initial
conditions are chosen appropriately, and a detailed understanding of the 
dynamical origin of the breaking point is missing and may profit from knowledge
about its $\sqrt{s_{_{NN}}}$-dependence.

Third, mass-ordered $v_2(p_T)$ at mid-rapidity are predicted in transport 
approaches~\cite{Burau:2004ev,Lu:2006qn} as well as in hydrodynamical 
models~\cite{Teaney:2001av,Kolb:2003dz}, but details of the 
predicted mass hierarchy vary between models.
For instance, two different ideal-fluid dynamics approaches predict scaling laws
with different variables: the decoupling of all particle species from the same 
collective-flow field is argued in reference~\cite{Borghini:2005kd} to lead to 
a common $v_2(p_T/m)$ for all hadrons; while in the Buda--Lund framework~\cite{%
  Csanad:2005gv} a scaling of $v_2(p_T)$ with the square of the transverse 
rapidity $y_T\equiv\frac{1}{2}\ln[(m_T+p_T)/(m_T-p_T)]$ is predicted. 
So, while mass ordering of $v_2(p_T)$ for light hadrons is expected to persist 
qualitatively, its quantitative manifestation may help to differentiate between 
models of the collision dynamics. 
 
Moreover, at the LHC, the elliptic-flow parameters $v_2(p_T)$ of $D$- and 
$B$-mesons will provide yet another test of the mass-ordering of $v_2$. 
Quantitative comparison to predictions of dynamical approaches --- be it fluid 
dynamics, a Langevin description, or a transport model --- have been argued to
give insight on the possible thermalization~\cite{Moore:2004tg} of heavy quarks, 
in particular their mean free path~\cite{Zhang:2005ni}, and their hadronization 
mechanism~\cite{Lin:2003jy,vanHees:2005wb}.
Quite generally, heavy quarks in equilibrium have a larger $v_2(p_T)$ than 
non-equilibrated ones; and the corresponding mesons have larger elliptic-flow 
values if they form through coalescence (involving a light quark) than if they 
come from heavy-quark fragmentation.

Mass-ordering persists essentially up to the breaking point in the $p_T$-shape.
Above this point and up to $\approx 5-6\mbox{ GeV}/c$, a quark-counting rule was
observed at RHIC for light hadrons, defining a region of intermediate transverse
momenta. 
(New $n_q$-scaling rules at low transverse momentum were also recently reported 
at RHIC~\cite{Adare:2006ti,Abelev:2007qg}, which to our knowledge have no 
theoretical explanation in the existing literature.) 
The persistence of such rules at LHC would constrain dynamical models, 
especially those covering both the low- and intermediate-$p_T$ regions. 
 
We now turn to the centrality dependence of $v_2$. 
It has been suggested to classify finite impact-parameter collisions for nuclei 
of different size in terms of the surface $S$ of the transverse overlap region 
and the eccentricity $\epsilon$ of this surface. 
(We note that fluctuations~\cite{Alver:2007qw,Bhalerao:2006tp} and uncertainties
in the initial conditions~\cite{Adil:2006gm,Lappi:2006xc} make the specification
of $\epsilon$ and $S$ somewhat model-dependent.) 
One observes that data of $v_2(y=0)/\epsilon$ from AGS to RHIC energy show an 
apparently universal, linear dependence if plotted against 
$(1/S)\,{\rm d}N^{\rm ch}/{\rm d}y$ (see e.g.\ figure~25 in reference~\cite{Alt:2003ab} 
or figure~15 in reference~\cite{Roland:2005ei}).
Modelling the matter produced in heavy ion collisions in terms of a perfect 
fluid, one finds that the above-mentioned linear dependence levels off above 
values of $(1/S)\,{\rm d}N^{\rm ch}/{\rm d}y$ corresponding to central RHIC 
energies. 
Thus, as in our discussion of several other classes of measurements, a naive 
extrapolation of these data on centrality from RHIC to LHC is at odds with a 
naive extension of the main dynamical explanation advocated to underlie 
elliptic flow $v_2$. 
On the other hand, there are models which may account for a further increase of 
$v_2/\epsilon$ due to non-equilibrium phenomena in the initial state~\cite{%
  Bhalerao:2005mm}, or due to a change in the relative contributions of hadronic
and partonic rescattering effects as a function of $\sqrt{s_{_{NN}}}$~\cite{%
  Hirano:2007gc}, or due to a significant change in initial conditions~\cite{%
  Hirano:2005xf}.

The second most studied flow harmonic is $v_1$, ``directed flow'', which 
quantifies the average momentum acquired by the particles along the 
impact-parameter direction.
Two generic trends can be identified in existing data:
\begin{enumerate}
\item \underline{The $p_T$-integrated $v_1(y)$ is linear around mid-rapidity.}\\
At mid-rapidity, $v_1$ vanishes by symmetry in collisions between 
identical nuclei. A linear increase of $v_1(y)$ around $y=0$ is observed 
across AGS and SPS energies. The slope ${\rm d}v_1/{\rm d}y$ decreases with 
increasing beam energy.
\item The $p_T$-integrated $v_1(\eta)$ shows \underline{extended 
longitudinal scaling~\cite{Back:2005pc}}.\\
Above SPS energy, one finds that  
$v_1(\eta)$ is positive in the ``projectile'' ($\eta>0$) fragmentation region%
\footnote{This is a choice, inherited from fixed-target studies at lower 
  energies, where the bounce of the projectile off the target is taken to define
  the positive direction. 
  The absolute sign of $v_1$ is not measurable, only its changes in sign are.},
then becomes negative for $\eta\lesssim y_{\rm beam}$, reaches a minimum for 
$-2\lesssim\eta-y_{\rm beam}\lesssim -1$, and increases. 
\end{enumerate}
The requirement that $v_1$ be zero at midrapidity implies either a 
breakdown of the longitudinal-scaling property, or that $v_1(\eta)$ vanishes in 
an extended region around $y=0$ (so that the slope ${\rm d}v_1/{\rm d}y$ is also
zero), the size of which increases with beam energy. 
In either case, the $p_T$-integrated $v_1(y)$ at LHC will be smaller in absolute
value than 0.01 up to rapidities $y\approx 4-5$.
A deviation from this expectation would indicate a physics effect that 
manifestly breaks Bjorken boost-invariance and which becomes more pronounced at 
higher $\sqrt{s_{_{NN}}}$. 
We are not aware of any suggestion of such an effect at these energies. 
Thus, testing a non-trivial dependence of $v_1$ at LHC and extending the RHIC 
systematics is likely to require measurements at far forward rapidity, which are
experimentally challenging. 

We finally comment on the fourth anisotropic-flow harmonic $v_4$.
It has been argued that at given transverse momentum 
and rapidity, $v_4(p_T,y)\geq\frac{1}{2} v_2(p_T,y)^2$~\cite{Borghini:2005kd,%
  Bhalerao:2005mm}, the lower bound being attained if and only if the matter 
expands like a perfect fluid at the time when anisotropic flow develops.
At RHIC, one has reported $v_4/v_2^2 \simeq 1.2$~\cite{Adams:2003zg}.
If one takes this number as an indication of incomplete equilibration, and if one assumes
that equilibration mechanisms are more efficient at the LHC, one predicts
$0.5 < v_4/v_2^2 < 1.2$ at the LHC. Without such assumptions, one still observes that 
both RHIC data and model calculations
lead us to expect a value of $v_4/v_2^2$ of order unity.

\section{Femtoscopy}
\label{s:femtoscopy}

\begin{figure}[t]
\centerline{\includegraphics*[width=10cm]{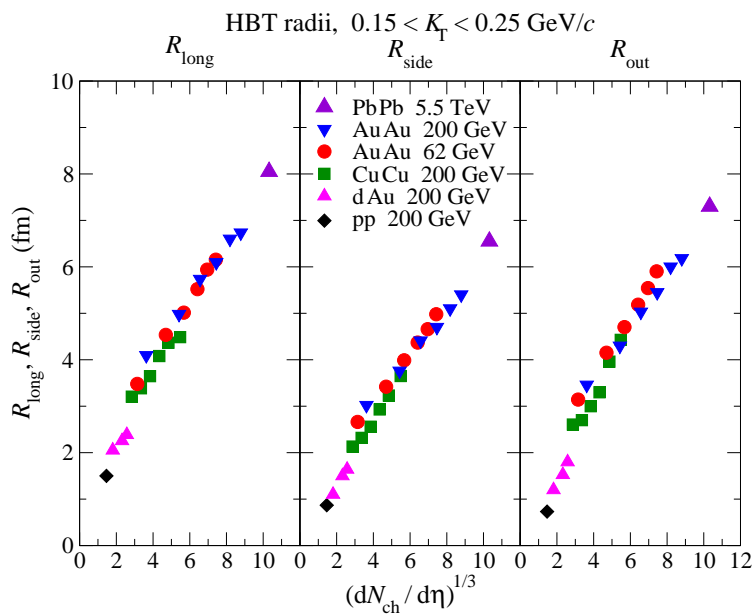}}
\caption{The diagonal HBT radius parameters $R_{\rm l}$, $R_{\rm s}$ and $R_{\rm o}$
  at mid-rapidity plotted versus event multiplicity for different centre-of-mass
  energies, including an extrapolation to LHC. 
  Data from the STAR collaboration taken from reference~\cite{Chajecki:2006sf}. 
  Except for the 200~GeV Au-Au results, all data are preliminary, but consistent
  with previously observed trends~\cite{Lisa:2005dd}.}
\label{fig:HBT}
\end{figure}

The header {\it femtoscopy\/} summarizes a class of measurements that give 
access to the spatio-temporal extension and collective dynamics of the matter 
produced in heavy-ion collisions~\cite{Lisa:2005dd}.
This includes in particular identical two-particle momentum correlations 
$C({\bf K}, {\bf q})$ of relative pair momentum ${\bf q}$ and average pair 
momentum ${\bf K}$. 
These are often analysed in the Bertsch-Pratt parametrization
\begin{equation}
\label{Bertsch-Pratt}
	C({\bf K}, {\bf q}) = 1+ \lambda \exp \left[ 
	\sum_{i,j={\rm o,s,l}} R_{ij}({\bf K}_{\perp},K_L)\, q_i\, q_j \right].
\end{equation}
Here,  the indices $i,j$ label a Cartesian coordinate system with axes pointing 
along the {\it longitudinal\/} beam direction ({\it longitudinal\/} or ``l''), 
parallel to the transverse pair momentum ${\bf K}_\perp$ ({\it out\/} or ``o'') and
the remaining ({\it side\/} or ``s'') one. 
The radius parameters $R_{ij}({\bf K}_{\perp},K_L)$ combine information about the 
spatial and temporal extension of the particle-emitting source at freeze-out.
They do not measure the extension of the entire collision region, but they 
measure the generally smaller ``homogeneity regions'', i.e. the part of the source
radiating particle pairs with pair momentum $({\bf K}_{\perp},K_L)$. 
Data of these HBT radii, taken at the SPS and RHIC, display several generic 
trends:
\begin{enumerate}
\item \underline{Almost linear scaling with 
  $\left({\rm d}N_{\rm ch}/{\rm d}\eta\right)^{1/3}$.}\\
  As seen in figure~\ref{fig:HBT}, the diagonal radius parameters in the out-, 
  side-, and longitudinal directions scale approximately linearly with the third
  root of the charged particle multiplicity per unit rapidity. 
  The entire $\sqrt{s_{_{NN}}}$-dependence appears to arise via the dependence on
  $({\rm d}N_{\rm ch}/{\rm d}\eta)^{1/3}$.
\item \underline{Universal $m_T$-dependence.}\\
  Values for the HBT radius parameters of different particle species fall on a 
  universal curve if plotted versus transverse mass, $m_T=\sqrt{m^2 + K_T^2}$.
  In particular, the ansatz
  \begin{equation}
	R_i(K_T) \simeq c_i  \frac{1}{m_T^{\alpha_i}}\, ,
	\label{eq4}
  \end{equation}
  provides a fair description of the universal $m_T$-dependence. 
  Here, the parameters $\alpha_i$ and $c_i$ take the same universal values for 
  all hadron species. 
  Fit values of $\alpha_i$ are $\sim 0.5$, albeit with large variations.
\item \underline{$R_{\rm o}^2 \simeq R_{\rm s}^2$.}\\
  The three diagonal radius parameters differ in size, but this difference is 
  small compared to their absolute value. 
  In particular, $R_{\rm o}^2 \simeq R_{\rm s}^2$.
\end{enumerate}
%

The extrapolation of the linear-scaling trend (i) from RHIC to the LHC leads to 
values for the HBT radius parameters that exceed significantly those measured 
previously, see figure~\ref{fig:HBT}. 
For instance, one finds 
$R_{\rm l}(150\mbox{ MeV}\!<\! K_T \!<\! 250\mbox{ MeV}) \simeq 8\mbox{ fm}$ for
${\rm d}N_{\rm ch}/{\rm d}\eta = 1100$, and even larger values for larger event 
multiplicities. 
For central collisions at mid-rapidity, the off-diagonal radius parameters 
$R_{ij}$ vanish by symmetry, and the product $V=R_{\rm l}\,R_{\rm s}\,R_{\rm o}$ 
provides a working definition of the spatial volume of the homogeneity region. 
The linear increase of all three HBT radii with 
$({\rm d}N_{\rm ch}/{\rm d}\eta)^{1/3}$ is then consistent with the statement, that
hadrons freeze-out from the collision system at an universal phase-space 
density~\cite{Ferenc:1999ku}. 
Changes in hadrochemical composition, transverse flow or temperature gradients 
may lead to deviations from this freeze-out criterion at universal phase-space 
density~\cite{Sinyukov:2002if}.
In particular, an increase of the fraction of baryons ($p$ + $\bar{p}$ + higher 
resonances) at fixed density leads to an increase of HBT radius parameters, 
since baryonic cross sections are larger than mesonic ones and thus delay 
freeze-out~\cite{Adamova:2002ff}.
In contrast, larger flow or temperature gradients tend to decrease the HBT 
radius parameters, since they narrow the spatial extension within which 
identical particle pairs show significant quantum-mechanical interference. 
The linear extrapolation $R_i \propto ({\rm d}N_{\rm ch}/{\rm d}\eta)^{1/3}$ 
provides an agnostic baseline on top of which dynamical changes may be 
established. 

In models with flow-dominated freeze-out scenarios, the $m_T$-dependence of the 
transverse HBT radii steepens as flow increases and/or temperature decreases. 
However, the numerical significance of this effect is model-dependent and other 
factors (such as effects from resonance decay contributions or the opacity of 
the produced matter) may play a role as well. 
Even if one assumes that transverse flow effects increase with $\sqrt{s_{_{NN}}}$ 
from RHIC to LHC, the question whether the $m_T$-dependence of transverse HBT 
radii steepens, appears to be too subtle to have a robust, model-independent 
answer. 
Moreover, in model studies, transverse flow may manifest itself in a 
$K_T$-dependence of transverse HBT radii, which cannot be absorbed in an 
$m_T$-dependence. 
We are not aware of sufficiently generic LHC predictions for this interesting 
class of measurements. 
As an aside, we note that for establishing the trend displayed in 
figure~\ref{fig:HBT}, data for different centralities and centre-of-mass energies 
should be compared at the {\it same} transverse momentum --- otherwise, the 
strong $m_T$-dependence (\ref{eq4}) may mask the apparently universal 
dependence on $({\rm d}N_{\rm ch}/{\rm d}\eta)^{1/3}$.

Under relatively mild model assumptions, the difference 
$R_{\rm o}^2(K_T) - R_{\rm s}^2(K_T)$ can be related to the lifetime of the 
particle-emitting source. 
The smallness of this quantity has been dubbed puzzling, since many models of 
the source dynamics predicted large lifetimes (for instance as a consequence of 
a first order phase transition or rapid crossover)~\cite{Rischke:1996em}. 
Also, fluid dynamics strongly over-predicts $R_{\rm o}^2 - R_{\rm s}^2$, if it is not
supplemented by extended hadronic scattering mechanisms~\cite{Soff:2000eh}, or by
other significant modifications of the final state~\cite{Miller:2005ji}.
In the energy range from RHIC to LHC, no mechanism is expected to set in, which 
could modify the value $R_{\rm o}^2(K_T) - R_{\rm s}^2(K_T)$ significantly. 

The analysis of femtoscopic information at LHC is not limited to the three 
generic features listed above. 
It includes the analysis of HBT radii with respect to the reaction plane~\cite{Wiedemann:1999qn}, the 
physics hidden in the intercept parameter $\lambda$ of 
equation~(\ref{Bertsch-Pratt}), and a rapidly growing field of non-identical particle
correlations. 
So far, however, data on these classes of observables are too scarce to serve 
as a robust baseline for extrapolations to the LHC, and we refer to a recent
review article for further discussions~\cite{Lisa:2005dd}. 

\section{Single inclusive high-pt spectra in $A$-$A$ and $pA$ collisions}
\label{s:high_pT}

The study of single inclusive hadron spectra at high transverse momentum has lead to
some of the major discoveries of the RHIC heavy ion program~\cite{Jacobs:2004qv}. 
At the SPS, kinematic
constraints limit the analysis of transverse momentum spectra in practice to $p_T \leq 3 - 4$~GeV/$c$. 
In contrast, the RHIC program studied at higher centre-of-mass energy 
various hadron species in a range up to $p_T \leq 10 - 20$~GeV/$c$, where perturbative 
production mechanisms are known to account for the single inclusive spectra in 
hadronic collisions. At the LHC, the transverse phase space accessible for such measurements 
increases by another factor $\sim 10$. The limited kinematic reach of the SPS prompts
us in the following sections to base extrapolations to the LHC on RHIC data only.

The nuclear modification factor $R^h_{AB}$ characterizes how the production of
a hadron $h$ in a nucleus-nucleus collisions $A$-$B$ differs from its production in an 
equivalent number of proton-proton collisions, 
\begin{equation}
R^h_{AB}(p_T,\eta,{\rm centrality})= {  {{\rm d}N^{AB\to h}_{\rm medium}\over 
{\rm d}p_T\, {\rm d}\eta} \over
\langle N^{AB}_{\rm coll}\rangle {{\rm d}N^{pp\to h}_{\rm vacuum}\over 
{\rm d}p_T\,{\rm d}\eta}}\, .
\label{eq5}
\end{equation}
Here,  $\langle N^{AB}_{\rm coll}\rangle$ is the average number of inelastic nucleon-nucleon 
collisions in a given centrality class. This number is typically determined in a Glauber-type 
calculation. The nuclear modification factor depends in general
on the transverse momentum $p_T$ and pseudo-rapidity $\eta$ of the particle, the particle
identity $h$, the centrality of the collision and the orientation of the particle trajectory with respect
to the reaction plane (which is often averaged over). In the absence of medium effects, 
$R^h_{AB} = 1$. 
%
\subsection{The nuclear modification factor at mid-rapidity}
RHIC data on $R_{AA}$ show the following generic features:
\begin{enumerate}
	\item \underline{Characteristic centrality dependence of $R_{AA}$ and $R_{\rm dAu}$.}\\
	For the most peripheral centrality bin, the nuclear modification factors measured
	at RHIC are consistent with the absence of medium-effects in both nucleus-nucleus
	($R_{AA} \sim 1$) and deuterium-nucleus ($R_{\rm dAu} \sim 1$) 
     collisions~\cite{Arsene:2003yk,Adler:2003ii,Back:2003ns,Adams:2003im}. With increasing
	centrality, $R_{AA}$ decreases monotonically. In d-Au collisions, the opposite centrality
	dependence is observed with  maximal values $R_{\rm dAu} \sim 1.5$ around $p_T = 3-5$~GeV/$c$
	in the most central bin.
  Accordingly, one observes that $R_{AA}$ depends on the azimuth with respect to 
  the reaction plane, with a smaller $R_{AA}$ out-of-plane~\cite{Adler:2006bw} 
  --- equivalently, $v_2(p_T)$ is positive at high $p_T$~\cite{Adams:2004bi}.
	\item \underline{Strong and apparently $p_T$-independent suppression of $R_{AA}$ at high $p_T$~.}\\
	In $\sqrt{s_{_{NN}}}= 200$ GeV, 5-10\% central Au-Au collisions at mid-rapidity, one observes a 
	suppression of high-$p_T$ single inclusive hadron yields by a factor $\sim 5$, corresponding to 
	$R^h_{AuAu}(p_T) \simeq 0.2$ for $p_T \geq 5-7$~GeV/$c$. Within experimental errors, this 
	suppression is $p_T$-independent for higher transverse momenta in all centrality 
     bins~\cite{Adams:2003kv,Back:2004ra,Adler:2006hu}.
	\item \underline{Independence of $R_{AA}$ on hadron identity.}\\
	For transverse momenta $p_T\geq 5-7$~GeV/$c$, all identified (light-flavoured) hadron spectra
	show a quantitatively comparable degree of suppression. There is no particle-species dependence
	of the suppression pattern at high $p_T$. 
	\item\underline{The photon spectrum is consistent with perturbative expectations.}\\
     For single inclusive photon spectra, the nuclear modification factor shows mild
     deviations from $R_{\rm AuAu}^\gamma = 1$~\cite{Isobe:2007ku}. 
  Within errors, these are consistent with perturbative predictions taking into 
  account the nuclear modifications of parton distribution functions (mainly 
  the isospin difference between protons and nuclei)~\cite{Arleo:2006xb}. 
     \item\underline{$R_{AA}$ shows a characteristic baryon-meson difference at intermediate $p_T$.}\\
     At intermediate $p_T$, $3\mbox{ GeV}/c < p_T < 6\mbox{ GeV}/c$ say, the nuclear modification 
     factor for mesons is smaller than the one for baryons~\cite{Abelev:2006jr}. Within 
     experimental uncertainties and
     irrespective of the hadron mass, all identified meson spectra show a similar degree of 
     nuclear suppression, and so do all identified baryon spectra. 
     \end{enumerate}
For hadronic collisions, the perturbative QCD factorized formalism can account systematically
for single inclusive hadron spectra at sufficiently high transverse momentum, by convoluting (``incoming'')
parton distribution functions, with hard, partonic scattering matrix elements, and with
(``outgoing'') parton fragmentation functions. In nucleus-nucleus collisions, one aims at
identifying the leading medium-length enhanced nuclear effects which modify this factorized
formalism~\cite{Qiu:2005ki}. In this context, the notions ``ingoing'' and ``outgoing'' become physically relevant, since the hard production process is placed within the  spatio-temporal geometry 
of a nuclear collision. The zeroth order question is whether the dominant medium modification
is accumulated during the incoming or outgoing stage of its prolonged interaction with the 
medium. Experimentally, this can be addressed by systematically varying the final state effects; 
for instance by varying the outgoing in-medium path length via centrality measurements, or 
by switching off final state effects by comparing $A$-$A$ collisions with $hA$ collisions. 
Theory addresses these dependencies in model studies,
which supplement the perturbative QCD factorization approach to single inclusive hadron
spectra with medium-modifications in the initial and final state, taking the spatio-temporal
distribution of matter during the $A$-$A$ collision into account~\cite{Baier:2000mf,Kovner:2003zj,Gyulassy:2003mc}.

RHIC data prove that high-$p_T$ hadron suppression is predominantly a final state effect by 
establishing that the suppression is not seen in d-Au and that it increases
in $A$-$A$ with increasing centrality and thus with increasing in-medium path length
in the final state [point (i)]. Moreover, the independence of $R_{AA}^h$ on hadron identity
[point (ii)] at high $p_T$ gives support to the picture that the final state medium modification of 
parton fragmentation is of partonic nature, i.e. that it occurs prior to the onset of hadronization.
This is so, since hadronic states would present absorption cross sections which can be expected
to differ significantly with hadron identity, and should thus lead to a hadron-specific splitting
of the nuclear suppression factor $R_{AA}^h$, which is not observed above $p_T \geq 5-7$~GeV/$c$. 
In addition, the single inclusive photon spectrum indicates that initial state effects  at high $p_T$ 
are small and that they may be accounted for by nuclear 
modified parton distributions [point (iv)]. From these arguments, one concludes that the 
suppression of high-$p_T$ single inclusive hadron spectra in nucleus-nucleus collisions
is due to a {\it partonic, medium-length dependent final state effect}. 

From an agnostic point of view, one wonders whether the generic features (i)-(v) persist at
the 30 times higher $\sqrt{s_{_{NN}}}$ explored at the LHC, and how these features evolve
in the much wider transverse momentum range accessible at the LHC. To discuss these
questions, we first turn in section~\ref{s:en-loss@LHC1} to the extrapolation of models that describe 
the generic features observed in RHIC data. Then, we provide in section~\ref{s:en-loss@LHC2} a list of
effects that may become important at the LHC and could lead to characteristic
deviations from the generic features observed at RHIC.  

\subsubsection{Extrapolations of parton energy loss models to the LHC.}
\label{s:en-loss@LHC1}

The generic features (i)-(iv) are naturally 
accommodated in models which supplement the perturbative QCD factorization approach 
(using nuclear parton distribution functions) with a mechanism which degrades the energy of 
the leading outgoing partonic fragment due to its propagation in matter. 
Two classes of mechanisms have been explored in Feynman
diagrammatic detail: collisional~\cite{Bjorken:1982tu} and radiative parton energy 
loss~\cite{Gyulassy:1993hr,Baier:1996sk,Zakharov:1997uu,Wiedemann:2000za,Gyulassy:2000er,Wang:2001if}.
Radiative energy loss, that is the medium-enhanced splitting of the energetic parton,
is the dominant mechanism at high $p_T$~\cite{Baier:2000mf,Kovner:2003zj,Gyulassy:2003mc}. 
Up to which $p_T$
subleading collisional effects are numerically significant and how they could be 
disentangled from radiative ones is a matter of ongoing 
debate\cite{Djordjevic:2006tw,Adil:2006ei,Wang:2006qr}. In the following, we limit
our considerations to radiative energy loss, which gives a fair description of
RHIC data above $p_T > 7$~GeV/$c$~\cite{Baier:2001yt,Vitev:2002pf,Dainese:2004te,Eskola:2004cr}. 
This may provide a baseline on top of which collisional
contributions can be established.

In radiative parton energy loss models, only one medium-dependent model parameter enters,
the so-called jet quenching parameter $\hat{q}(\tau)$ (or a reparametrization of it), which  
depends on the time $\tau$ after the collision. In model studies, it is often expressed in terms
of the energy density $\epsilon(\tau)$~\cite{Baier:1996sk}
\begin{equation}
	\hat{q}(\tau) = c\, \epsilon^{3/4}(\tau)\, ,
	 \label{eq6}
\end{equation}
where $c$ is assumed to be a time- and temperature-independent constant. An estimate
based on perturbatively weak interactions between the hard parton and the medium gives
$c \approx 2$~\cite{Baier:2002tc}. In contrast, fitting parton energy loss models to RHIC data, 
several groups
found much larger values, $c \geq 8$~\cite{Eskola:2004cr,Dainese:2004te,Renk:2006sx}. 
Assuming a 1-dimensional Bjorken expansion of
the produced matter with $\hat{q}(\tau) = \hat{q}_0 \tau_0/\tau$, this translates into a value\footnote{Since medium-induced gluon energy radiation
in an expanding medium depends on the line-average $\bar{\hat{q}} = \frac{2}{L^2} \int_{0}^{L} {\rm d}\tau\,\tau\, \hat{q}(\tau)$~\cite{Salgado:2002cd}, several model studies quote $\bar{\hat{q}}$ for an 
  average in-medium path length~\cite{Eskola:2004cr,Dainese:2004te}. However, 
  $\bar{\hat{q}}$ depends strongly on $L$, which differs for each 
  parton. This introduces additional uncertainties. To bypass these problems, we quote here
  the value $c$, or equivalently $\hat{q}(\tau\!=\! 1\!\mbox{ fm}/c)$ which can be unambiguously extracted from all studies quoting $\bar{\hat{q}}$.
}
\begin{equation}
	\hat{q}(\tau = 1\!\mbox{ fm}/c) \geq 4 \mbox{ GeV}^2/\mbox{fm}\, .
	\label{eq7}
\end{equation}
The precise value of $\hat{q}$ consistent with RHIC data is currently debated, 
but it is generally thought that $\hat{q}$ is significantly larger than the 
perturbative estimate  $c = 2$ in reference~\cite{Baier:2002tc}.

The jet quenching parameter $\hat{q}$ has a rigorous field theoretical definition in terms of the 
short-distance behaviour of the target expectation value of a light-like Wilson 
loop~\cite{Liu:2006ug}. For a class
of non-Abelian thermal gauge field theories, which are known to have a gravity dual, 
non-perturbative evaluations of $\hat{q}$ have established recently that the seemingly
small scale of an initial temperature of $T(\tau\!=\! 1\!\mbox{ fm}/c) \simeq 300$ MeV does indeed 
give rise to a jet quenching parameter $\hat{q}(\tau\!=\! 1\!\mbox{ fm}/c)$ numerically consistent with
the apparently large lower bound of equation~(\ref{eq7})~\cite{Liu:2006ug}. These studies also show 
that the ratio of jet quenching parameters $\hat{q}$ of different thermal field theories is determined 
by the square root of the ratio of their entropy densities~\cite{Liu:2006he}. 
In earlier studies, $\hat{q}$ was taken
to be proportional to the event multiplicity, and large values up to $\overline{\hat{q}}_{\rm LHC}
\simeq 7\, \overline{\hat{q}}_{\rm RHIC}$ were explored~\cite{Eskola:2004cr,Dainese:2004te}. 
On the other hand, the multiplicity extrapolations to the LHC shown in figure~\ref{fig1} would
indicate a much smaller value of $\overline{\hat{q}}_{\rm LHC}$, and --- given that a
multiplicity scaling of $\hat{q}$ is an additional model assumption --- a very mild increase of say
$\hat{q}_{\rm LHC} \simeq 1.25\, \hat{q}_{\rm RHIC} $ is conceivable. 
Since $\hat{q}$ is the only medium-sensitive parameter in a class of model studies, once its value is fixed we can detail the predictions
of these energy loss models for the LHC:
\begin{enumerate}
	\item \underline{Centrality dependence.}\\
	Parton energy loss models that implement final state effects only, predict the absence of
	nuclear suppression in the most peripheral collisions. As a consequence, one expects
	that the centrality dependence of $R_{\rm PbPb}$ at the LHC parallels the one observed 
	at RHIC. 
	\item \underline{$p_T$-dependence of $R_{\rm PbPb}$ at the LHC.}\\
	At RHIC energies, the slope of the partonic $p_T$-spectrum gradually steepens
	as one moves from $p_T \sim 10$~GeV/$c$ to the absolute kinematic boundary
	$p_T = 100$~GeV/$c$. This implies that to obtain the same value of $R_{AA}$ at 
	higher $p_T$, a smaller fraction of parton energy loss is needed 
     (``trigger bias effect'', see reference~\cite{Baier:2001yt}).
	In contrast, at the LHC, the partonic $p_T$-spectrum will show almost the same
	power-law over the entire range $10\mbox{ GeV}/c < p_T < 100\mbox{ GeV}/c$, since this $p_T$-range is far away
	from the kinematic boundary at the LHC. This implies that for a $p_T$-independent
	$R_{AA}(p_T)$, one requires a constant, $p_T$-independent fractional energy loss,
	not predicted in current energy loss models~\cite{Eskola:2004cr}. For a mild 
     increase of e.g. $\hat{q}_{\rm LHC} \simeq 1.25\, \hat{q}_{\rm RHIC} $,
	models thus indicate that for central collisions,
	$R_{\rm PbPb}(10\mbox{ GeV}/c < p_T < 20\mbox{ GeV}/c)$ is \underline{the same or slightly larger} 
	(by up to $\sim 0.1$) than at RHIC, and that  $R_{\rm PbPb}(p_T)$
\underline{increases gently} by $0.1-0.2$ from
	$p_T = 10$~GeV/$c$ to $p_T \sim 100$~GeV/$c$\footnote{In this kinematic range, the partonic spectrum at the LHC is gluon dominated,
	while it is strongly quark dominated at RHIC. This difference in composition decreases
	the nuclear modification factor at LHC, compared to the one at RHIC, and compensates
	partially, but not completely for the trigger bias effect.}.
	
	\item
	\underline{Dependence of $R_{\rm PbPb}$ on hadron identity.}\\
	Light-flavoured hadron spectra are expected to 
	show the same nuclear suppression independent of hadron species at sufficiently high 
	$p_T > p_T^{\rm pid}$. If 
	the particle species dependence at intermediate $p_T < p_T^{\rm pid}$ is due to
	a medium-dependent hadronization mechanism (such as proposed for instance in
	recombination models~\cite{Fries:2003kq,Molnar:2003ff,Fries:2003vb,Greco:2003xt,Hwa:2002tu}), 
     then $p_T^{\rm pid}(\sqrt{s_{_{NN}}})$ is expected to
	increase with $\sqrt{s_{_{NN}}}$ by up to $2-3$~GeV/$c$ from RHIC to the LHC~\cite{Fries:2003kq}. In contrast,
	if the scale $p_T^{\rm pid}$ is mainly set by the time dilation of the hadronization 
     time~\cite{Wiedemann:2004wp}, 
	implying that for $p_T > p_T^{\rm pid}$ hadronization occurs outside the medium, then
	one may expect that $p_T^{\rm pid}$ is $\sqrt{s_{_{NN}}}$-independent. Thus, the
	$\sqrt{s_{_{NN}}}$-dependence of $p_T^{\rm pid}$ may provide complementary information
     about the mechanism underlying the anomalous baryon-to-meson ratio at intermediate $p_T$.\\
	For heavy-flavoured hadrons, parton energy loss models predict a hierarchy in the
	nuclear suppression~\cite{Dokshitzer:2001zm,Armesto:2003jh,Djordjevic:2003zk,Zhang:2003wk}, 
    which can be characterized by ``heavy-to-light'' ratios of the 
	corresponding nuclear modification factors of heavy-flavoured over light-flavoured 
     hadrons~\cite{Armesto:2005iq}.
	For $D$-mesons, the charm mass is expected to be too small to contribute to a
	mass-dependent suppression above $p_T > 10$~GeV/$c$. However, since light-flavoured
	hadrons at the LHC are dominated by gluon parents, the heavy-to-light ratio of
	$D$-mesons is sensitive to the colour charge dependence of parton energy loss
	and is expected to exceed unity by up to a factor $\sim 1.5$ in an extended $p_T$-range.
	In contrast, for $B$-mesons, the mass effect is expected to put strong limits on
	medium-induced energy loss in an extended $p_T$-range. The resulting heavy-to-light 
	ratios are expected to be significantly larger than for $D$-mesons, reaching a factor 
	$\sim 2 - 4$, even for relatively low estimates of the value of $\hat{q}$~\cite{Armesto:2005iq}.
	\item
	\underline{Single inclusive photon spectra.}\\ 
     Based on a naive extrapolation from RHIC data, one may expect that  photon spectra
     deviate only mildly from $R_{AA}^{\gamma} \sim 1$ at the LHC. However, current
     parton energy loss models allow for mechanisms which may lead to significant 
     medium-modifications: The medium-induced photon bremsstrahlung of hard partons
     may enhance the photon yield at high $p_T$~\cite{Zakharov:2004bi};
     hard partons, which fragment into photons, may reduce the photon yield at 
     high $p_T$~\cite{Arleo:2004xj}.
     Both effects are of order $\alpha_{\rm em}$, and it is unknown to what extent they cancel
     each other and how they may vary as a function of $p_T^{\gamma}$. 
 \end{enumerate}
%
\begin{figure}[h]\epsfxsize=12.cm
\centerline{\epsfbox{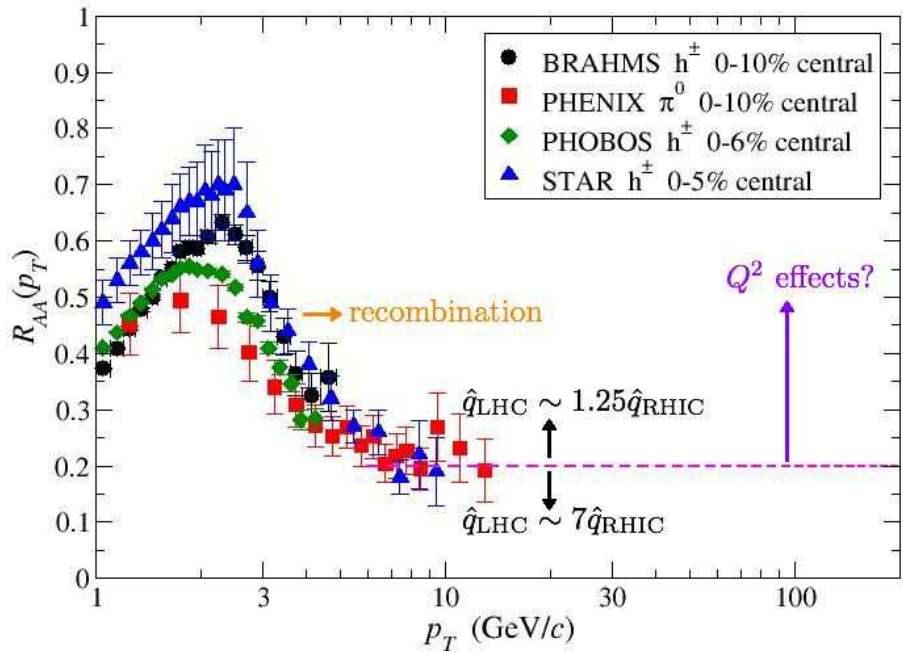}}
\caption{The nuclear modification factor $R_{AA}$ as a function of transverse momentum
at mid-rapidity. Data are for hadronic spectra measured at RHIC in $\sqrt{s_{_{NN}}}=200$~GeV 
Au-Au collisions. The dashed line is a straight continuation of the high-$p_T$ trend at RHIC,
reproduced by models of radiative parton energy loss. Arrows indicate qualitative tendencies
of how $R_{AA}$ may change at the LHC, see text for detailed discussion.  
}\label{fig6}
\end{figure}

\subsubsection{Testing effects not encoded in current parton energy loss models.}
\label{s:en-loss@LHC2}

We now turn to phenomena that are qualitatively novel in the sense that they are not
encoded in current parton energy loss models tested at RHIC, but may become important
at the LHC.
\begin{itemize}
 \item Is high-$p_T$ hadron suppression at mid-rapidity a final state effect at all $\sqrt{s_{_{NN}}}$?\\
  At RHIC, the absence of suppression in
 $R_{\rm dAu}$ showed conclusively, that initial state effects are unimportant 
 for $R_{AA}$~\cite{Adams:2003im,Adler:2003ii,Back:2003ns,Arsene:2003yk}. However, high-$p_T$ suppression in $pA$ has been predicted for sufficiently
 high $\sqrt{s_{_{NN}}}$ as a consequence of non-linear QCD evolution in the so-called geometric
 scaling window~\cite{Kharzeev:2002pc,Baier:2003hr,Albacete:2003iq,Kharzeev:2003wz,Jalilian-Marian:2003mf}. At present, the value of $\sqrt{s_{_{NN}}}$ at which
 non-linear evolution starts to become relevant for particle production at intermediate $p_T$ is unclear.
 If the onset of such non-linear evolution effects should lie below LHC energy, then this would be
 signalled by $R_{\rm pPb} < 1$ at LHC mid-rapidity. The enhancement $ R_{\rm dAu} > 1$ observed at
 RHIC  mid-rapidity would turn into a characteristic suppression as a function of $\sqrt{s_{_{NN}}}$. 
 Moreover, one expects in this case that the nuclear modification factor in Pb-Pb will be suppressed
 due to both, initial and final state effects. 
 \item Does $R_{\rm PbPb}(p_T)$ show indications of medium-dependent $Q^2$-evolution?\\
 QCD evolution 
 underlies the partonic fragmentation process in the perturbative regime,
 as well as the scale dependence of fragmentation functions. The question of how the
 medium affects the QCD scale evolution is difficult to address 
 theoretically, since medium effects are ``higher twist'' , i.e. subleading by powers of $Q^2$,
 although they may be nuclear enhanced by geometric factors 
 $\propto A^{1/3}$~\cite{Luo:1992fz,Guo:2000nz}. Qualitatively,
 however, one may expect that --- despite the time dilation of the parton fragmentation process
 in the target rest frame --- at sufficiently large $Q^2$, parton splitting occurs on length
 scales too short to be resolved by the medium, and short-distance contributions to
 parton fragmentation remain unmodified accordingly. It has been speculated that this may lead
 to a significant rise of $R_{AA}(p_T)$ in the range 
 $20\mbox{ GeV}/c < p_T < 100\mbox{ GeV}/c$~\cite{Borghini:2005em},
 but detailed model studies are still missing. More generally, the fact that LHC has access to
 a logarithmically wide $p_T$-range may provide novel opportunities to test the medium
 dependence of QCD-evolution.
 \item
 What is the dynamical mechanism underlying the nuclear modification of quarkonium?\\ 
 Data on $R_{AA}^{J/\Psi}(p_T,\eta)$ at RHIC are known up to less than $p_T < 5$~GeV/$c$. They
 indicate suppression at small $p_T$,
 $R_{AA}^{J/\Psi}(p_T\!\lesssim\! 1\,{\rm GeV}/c,\vert\eta\vert\!\lesssim\! 0.35) \simeq 0.3-0.4$, and possibly
 an increase of $R_{AA}^{J/\Psi}$ with $p_T$. This is qualitatively similar to $R_{AA}^{h}$
 of light hadrons. However, the physics invoked to account for the suppression
 of $J/\Psi$'s shows marked differences if compared to the high-$p_T$ suppression of 
 light hadrons. Quarkonium suppression arises from the fact that the attraction
 between heavy quarks and anti-quarks weakens with increasing temperature due to 
 dynamic screening effects~\cite{Matsui:1986dk}.  According to recent models, directly
 produced $J/\Psi$'s may not dissociate until well above the energy densities attained at 
 RHIC, but  the $\chi_c$'s and $\Psi'$'s, whose decay contributions are estimated to 
 $\sim 60 \%$ of the $J/\Psi$ yield in hadronic collisions, are expected to dissociate. 
 The observed value
 $R_{AA}^{J/\Psi}(p_T<1\, {\rm GeV}/c,\vert\eta\vert < 0.35) \simeq 0.3-0.4$ thus appears
 natural, if one assumes that all excited $c\bar{c}$ bound states are dissociated in the 
 medium~\cite{Karsch:2005nk,Satz:2005hx}.
\\
 At the LHC, one will have for the first time experimental access to significant rates of both
 bound charmonium and bottomium states. The lowest lying $b\bar{b}$-states are more
 tightly bound than the charmonium states, and one thus expects that they dissociate
 at higher temperature. Accordingly, in dissociation models one expects quite generally that
 the nuclear suppression of these bottomium yields is smaller or at most as large as that of
 the corresponding charmonium yields. A qualitatively opposite behaviour can be expected,
 if secondary production mechanisms, such as recombination, start playing a 
 role~\cite{Thews:2000rj}. These become
 more effective with the number of heavy quarks produced, and are thus more efficient in
 enhancing charmonium bound states. \\
 At high transverse momentum, $p_T > 5-7$~GeV/$c$, recombination effects are absent, and LHC
 will establish how the high-$p_T$ modification of the spectra of mesonic $Q\bar{Q}$ bound 
 states differs from that of heavy-light and light-light flavoured mesons. The following qualitative
 considerations illustrate the importance of formation time effects in this context: i) If high-$p_T$
 $J/\Psi$'s originate from gluon parents and if these gluons propagate over long distances
 before hadronizing, then the nuclear modification factor of $J/\Psi$'s is expected to be reduced
 due to gluon energy loss, and should take the same value as that for light-flavoured hadrons.
 ii) In contrast, if high-$p_T$ $J/\Psi$'s would originate from the fragmentation of $c$-quark
 parents and if the quark propagates over long distances prior to hadronizing, then the nuclear
 modification factor is expected to be reduced due to quark energy
 loss only, and should match the smaller reduction of heavy-light flavoured hadrons
 $R_{AA}^{J/\Psi} \simeq R_{AA}^D > R_{AA}^h$. In this case, one would also expect in
 $pp$ collisions open charm production associated with $J/\Psi$ production. iii) If in contrast the 
 formation time of the $J/\Psi$ is small compared to the extension of the collision region $\sim 10$ fm,
 then it is a $Q\bar{Q}$ bound state of velocity $v$ increasing with $p_T$, which propagates
 through the medium. 
For this case, non-perturbative calculations based on the AdS/CFT correspondence
suggest that the screening length $L_s$ of the $Q\bar{Q}$ potential at 
temperature $T$ shows a characteristic velocity scaling 
$L_s(v,T) \propto L_s(0,T)/\sqrt{\gamma}$~\cite{Liu:2006nn}. Model
 estimates indicate that depending on the binding energy of the $Q\bar{Q}$-state, it is 
 this velocity-dependent dissociation effect which may dominate in an extended intermediate
 $p_T$-regime. Only in this latter case will different bound states such as $J/\Psi$ and $\Psi'$
 show a different degree of suppression. 
 \item Are high-$p_T$ single inclusive photon spectra sensitive to final state medium effects?\\
 In hadronic collisions at the LHC, a significant fraction of the single inclusive photons arises from
 the fragmentation of quarks and (at next-to-leading order) gluons.  If the parent partons suffer 
 medium-induced 
 energy degradation prior to fragmenting into a photon~\cite{Arleo:2004xj}, then $R_{AA}^\gamma$ 
 is reduced and the elliptic flow of photons receives a positive 
 contribution~\cite{Fries:2002kt,Turbide:2005bz}. The strength of this effect may allow one to constrain the
 photon formation time but is difficult to estimate a priori. On the other hand, 
 the interaction of the produced partons with the medium can lead to additional bremsstrahlung
 photons~\cite{Zakharov:2004bi}. This effect increases $R_{AA}^\gamma$, and contributes to a 
 negative $v_2$ for photons~\cite{Turbide:2005bz}. To the best of our knowledge, these 
 expectations remain qualitative to date, but 
 the wider kinematic reach at LHC should help to disentangle them in the data.
 We finally note that similar modifications of the high-$p_T$ spectrum are not expected for
 $Z$-bosons, since all aspects of its production are local and unlikely to interfere with the
 typical length and momentum scales present in the produced QCD matter.\\
\end{itemize}

\subsection{The nuclear modification factor as a function of rapidity.}
\subsubsection{Evolution of the Cronin peak at RHIC.}
The notion ``Cronin peak'' refers to the enhancement of the nuclear modification factor $R_{hA}$
above unity at intermediate transverse moment, $2\mbox{ GeV}/c \lesssim p_T \lesssim 4$~GeV/$c$. 
At lower transverse momentum, the nuclear modification factor is 
generally suppressed $R_{hA}< 1$. We
hasten to remark that the notion ``Cronin peak'' is but a description and not an explanation of the
shape of $R_{hA}$. As discussed below, there is no completely satisfactory explanation of
the dynamical origin of the Cronin peak so far, but LHC is well-positioned to provide 
additional insight. 

The kinematically available $p_T$-range decreases with increasing rapidity, and data 
from RHIC are currently limited to $p_T < 5$~GeV/$c$ for $\vert\eta\vert > 2$. Within this limited
range, one observes the following  apparently generic trends in RHIC data~\cite{Debbe:2006kw}:
\begin{enumerate}
	\item \underline{Rapidity and $p_T$-dependence of $R_{\rm dAu}$.}\\
	At RHIC, $R_{\rm dAu}$ shows a typical Cronin enhancement for $\eta \leq 1$. This Cronin
     peak  in the range $2\mbox{ GeV}/c\leq p_T \leq 4$~GeV/$c$
     monotonically decreases in the deuteron fragmentation region.  One finds $R_{\rm dAu} \simeq 1$ 
     at  $\eta \simeq 1$, $R_{\rm dAu} \simeq 0.5$ at $\eta = 3$\cite{Arsene:2004ux,Back:2004bq,Adler:2004eh} and even smaller values for $\pi^0$'s at higher rapidity~\cite{Adams:2006uz}.  
     In contrast, in the nucleus fragmentation region,
     the ratio $R_{\rm cp}$ of central over an 
     equivalent number of peripheral collisions, which is closely related to $R_{\rm dAu}$,
     shows an enhancement~\cite{Adler:2004eh} above unity, which increases with
	increasing rapidity. 
	\item \underline{Centrality dependence of suppression pattern in d-Au.}\\
	At mid-rapidity, $R_{\rm cp}$ increases around the Cronin peak 
	for increasing centrality. In contrast, the centrality dependence at forward deuteron rapidity is 
     inverted~\cite{Adler:2004eh}:
	$R_{\rm dAu}$ decreases with increasing centrality for $\eta > 1$, and the centrality dependence
	at $\eta \simeq 1$ is negligible.
\end{enumerate}
In the context of RHIC data, the discussion of these phenomena has focused on two initial state 
effects, which we address now: multiple scattering and non-linear QCD evolution of the incoming 
parton distribution functions.

The Cronin peak is often thought of as the consequence of a multiple scattering picture, in which the 
partons in the deuteron wavefunction undergo multiple interactions in the target nucleus prior to
producing relatively high-$p_T$ hadrons. Incoherent multiple scattering of these incoming partons
leads to a transverse momentum broadening of the initial parton distribution, which translates into a correspondingly broadened single inclusive hadron spectrum. This can account for the 
observed Cronin peak at mid-rapidity~\cite{Accardi:2003jh,Vitev:2003xu}, though it is unclear 
whether it can account for the particle
species dependence of the effect. Moreover, at least in their current model implementations,
such multiple scattering models predict the persistence of the Cronin peak 
at forward rapidity, and they imply an increase of $p_T$-broadening with increasing
centrality at all rapidities~\cite{Vitev:2003xu}. This contradicts the generic trends seen in the 
RHIC data.

In studies of non-linear QCD evolution at small momentum fractions $x$, one generically
finds that the growth of unintegrated gluon distribution functions with $\ln 1/x$ (or with
$\ln \sqrt{s_{_{NN}}}$) is saturated up to a scale $p_T < Q_s(x)$, which grows with
$1/x$. Moreover, above this saturation scale, non-linear QCD evolution characteristically
changes the power-law in a wide geometric scaling window. Contact between these
findings and the phenomenology of d-Au collisions is made by  the following observations:
First, toward deuteron projectile rapidity, smaller momentum fractions $x_{\rm Au}$ 
of the nuclear parton distribution functions become relevant for particle production. Eventually,
$x_{\rm Au}$ will become small enough for non-linear QCD evolution to
be applicable. Second, convoluting non-linear evolved unintegrated nuclear parton distributions
schematically with hard processes, one finds that non-linear QCD evolution implies the
decrease of $R_{hA}$ with increasing 
$\ln 1/x_{\rm Au}$~\cite{Kharzeev:2002pc,Baier:2003hr,Albacete:2003iq,Kharzeev:2003wz,Jalilian-Marian:2003mf}. This provides a conceivable explanation for the rapidity dependence of $R_{\rm dAu}$.

Non-linear QCD evolution does not account for all trends seen in the data. In 
particular, the Cronin peak itself is not a dynamical consequence of non-linear QCD evolution,
it is just a conceivable initial condition, which is quickly washed out by the evolution.
So, if saturation physics is the correct explanation for the rapidity dependence of 
$R_{\rm dAu}$, then one knows that it is not applicable at RHIC mid-rapidity. 
Also, while particle species identified data on $R_{\rm dAu}$ are not accurate enough to
allow for decisive tests, the question whether a purely partonic explanation is sufficient
to account for $R_{\rm dAu}$ in the experimentally tested range remained open so far. 
Moreover, it is unclear on theoretical grounds whether saturation physics can be expected 
to apply for the relatively large values $x_{\rm Au} \geq 0.02$~\cite{Guzey:2004zp} which 
dominate forward particle production at RHIC.

Experiments at the LHC will allow us to compare the $\sqrt{s_{_{NN}}}$- and $\eta$-dependence. 
To illustrate that this may provide a decisive test for current models, let us consider
an alternative explanation of the rapidity dependence of $R_{\rm dAu}$ at RHIC, based on the
following picture:  Partons in the deuteron wave function undergo {\em inelastic\/} multiple scatterings 
on the nuclear target field. Hence, these partons split due to interactions with the target. Since 
splitting is very effective in energy degradation, this will deplete the hadron yield at forward rapidity 
but will enhance it at mid-rapidity. Also, the opposite centrality dependence at mid-rapidity and
forward rapidity [point (ii)] can be understood in this way. 
In this picture, the suppression in $R_{\rm dAu}$ 
increases strongly with increasing projectile rapidity, because the energy degradation 
occurs on top of an increasingly steeply falling spectrum at forward rapidity. Transverse momentum
broadening could still contribute to the Cronin peak at mid-rapidity, but would not be able
to overcome the reduction at forward rapidity. This picture accounts for the same rapidity 
dependence as saturation models. In contrast to saturation models, however, it implies that 
the Cronin peak will not disappear with increasing $\sqrt{s_{_{NN}}}$, but 
will persist at LHC energies. 

\subsubsection{Conceivable effects on the $\eta$-dependence of $R_{\rm pPb}$ and $R_{\rm PbPb}$ at the LHC.}

The example given above illustrates that comparing the $\sqrt{s_{_{NN}}}$- and $\eta$-dependence
at the LHC will provide a qualitatively novel test for the saturation physics interpretations of
measurements at RHIC. This is so, since an increase in both $\sqrt{s_{_{NN}}}$ or $\eta$
gives access to smaller momentum fractions $x$ in the parton distributions, and thus has 
similar implications in saturation models. At the LHC, the $\sqrt{s_{_{NN}}}$-dependence of measurements, 
in combination with their $\eta$-dependence, will become a tool to discriminate effects from 
small-$x$ QCD evolution from other  conceivable mechanisms. 

Focusing in the following solely on the nuclear modification factor, we now list phenomena which
may affect significantly the  $\eta$-dependence of $R_{\rm PbPb}$ at the LHC, and which may
be disentangled by studying the $\sqrt{s_{_{NN}}}$-dependence in a combination of
data from RHIC and LHC.  Our list starts with
conceivable final-state effects:
\begin{enumerate}
	\item \underline{Dependence of $R_{\rm PbPb}$ on ${\rm d}N_{\rm ch}/{\rm d}\eta$ for fixed centrality.}\\
	As discussed in section~\ref{s:en-loss@LHC1}, the quenching parameter $\hat{q}$ is expected to grow 
	monotonously with  ${\rm d}N_{\rm ch}/{\rm d}\eta$. So, the quenching parameter should be 
	smaller at forward rapidity. This effect contributes to an {\it increase\/}
	of $R_{\rm PbPb}$ with $\eta$. 
	\item \underline{The $\eta$-dependence of the partonic $p_T$-spectrum.}\\
	With increasing $\eta$, partonic $p_T$-spectra get steeper. This is a simple kinematic
	effect, present both at RHIC and at the LHC, but quantitatively different. As a consequence
	of this effect, the same amount of parton energy loss leads to a {\it decrease\/} of $R_{\rm PbPb}$ 
	with $\eta$ (trigger bias).
	\item \underline{Flow effects on $R_{\rm PbPb}$.}\\
	The initial parton, produced in a hard collision, needs not be produced within 
	longitudinal comoving matter. In case that it is not, there is a relative longitudinal velocity
	between the hard projectile and the medium, and energy loss is expected to be higher.
	This effect is likely to contributes to a {\it decrease\/}
	of $R_{\rm PbPb}$ with $\eta$~\cite{Salgado:2003rv,Armesto:2004pt}, though estimates
     of its magnitude vary widely~\cite{Baier:2006pt}.  
        
	\item \underline{Initial state effects.}\\
	As discussed above, the nuclear modification of parton distribution functions is expected
	to affect $R_{\rm PbPb}$. In particular, models based on non-linear small-$x$ evolution
	predict~\cite{Kharzeev:2002pc,Baier:2003hr,Albacete:2003iq,Kharzeev:2003wz,Jalilian-Marian:2003mf} an additional {\it decrease\/} of $R_{\rm PbPb}$ with $\eta$. These models can 
	be tested by comparing the $\sqrt{s_{_{NN}}}$- and $\eta$-dependences in $p$-Pb collisions.
\end{enumerate}

\section{``Jet-like'' particle correlations and jets}
\label{s:high_pT2}

The leading hadronic fragments of highly energetic parent partons, measured in single inclusive 
hadron spectra, are strongly modified at RHIC and they are expected to 
be strongly modified at the LHC. Any model of the dynamical mechanism underlying this
medium modification has implications for the entire parton fragmentation pattern, and that
is: jets and jet-like observables. Jet measurements in heavy ion collisions are sensitive to
how high-energy partons are attenuated in matter and how they
equilibrate kinetically and chemically. In turn, these medium-modifications of jet fragmentation
characterize properties of the produced medium.

In general, parton fragmentation leads to multiplicity distributions with broad variances. As a consequence, any particle trigger used to select jet-like observables will bias significantly
the fragmentation pattern. Even prior to invoking medium effects, such biases have dramatic
consequences: In a typical single inclusive hadron spectrum (i.e. single particle trigger) at
$p_T > 20$~GeV/$c$, the hadrons will typically carry on average $\sim 3/4$ of the energy of their
parent partons. In contrast, the leading hadron in a 100~GeV/$c$ jet, initiated by a light
parton, carries typically only $\sim 1/4$ of the jet energy, simply because this jet fragmentation pattern
is not biased by a single particle trigger. In the presence of a medium, additional ``trigger biases'' may arise.
For instance, in the presence of strong final state energy loss, a high-$p_T$ particle trigger
will select particles produced mainly at the outskirts of the nuclear overlap region. The parent
partons of these hadrons 
have had a particularly small in-medium path length and thus suffered particularly little
parton energy loss (surface bias)~\cite{Muller:2002fa}. Also, a high-$p_T$ particle trigger 
will prefer events in
which the initial state $p_T$-broadening effects move the dijet invariant mass towards the
trigger. Thirdly, triggering on a high-energy hadron or requiring a jet can lead to structures
in the distribution of soft ``background'' particles, which are typically counted towards
the medium, but which are related to the trigger and would not be found in minimum bias
events. These general considerations prompt us to distinguish in the following discussion
between ``true'' jets, jet-like particle correlations and soft structures causally related to high-$p_T$
triggers.

\subsection{The medium-modification of ``true'' jets}

``True'' jet measurements, that is measurements of the {\it entire} fragmentation pattern of high-$E_T$
parent partons, have not been performed in heavy ion collisions so far. In the context of
RHIC data, ``jet quenching'' refers to the suppression of single inclusive
hadron spectra and high-$p_T$ particle correlations. Yet, measurements at RHIC, as
well as models of parton energy loss, give rise to a set of general expectations for ``true'' jet
measurements in heavy ion collisions:

\begin{enumerate}
	\item \underline{Longitudinal jet multiplicity distributions soften.}\\
	Parton energy loss, combined with energy-momentum conservation implies that
	the energy lost by the leading parton or hadron in the parton shower reappears
	in additional multiplicity of softer fragments. The entire longitudinal jet
	multiplicity distribution is expected to soften, and the total jet multiplicity is
     expected to increase, see e.g. references~\cite{Pal:2003zf,Borghini:2005em}.
	\item \underline{Transverse jet multiplicity distributions broaden.}\\
	Essentially all models of parton energy loss assume a significant transverse momentum transfer 
	from the medium to the jet projectile. As a consequence, 
	parton energy loss is generally thought to be accompanied by a 
	broadening of the jet fragmentation pattern in the plane orthogonal to the jet 
	axis~\cite{Salgado:2003rv,Armesto:2004pt,Armesto:2004vz,Renk:2005ta}. In case that the
	momentum transfer from the medium is asymmetric, for instance since the parton
	is embedded in a collective flow field, this jet broadening may show characteristic
	asymmetries~\cite{Armesto:2004pt,Armesto:2004vz}.
	\item \underline{The hadrochemical composition of jet fragments may be modified.}\\
	To date, most studies of jet medium-modifications focus on the longitudinal and
	transverse energy and multiplicity distributions. However, in current models of parton energy 
	loss, the medium couples to the parton shower via gluon exchange, and thus alters  the
	colour flow in the shower. This may be expected to affect the hadrochemical  composition
	of the jet. Also, in principle, other quantum numbers such as baryon number or flavour
	may be exchanged between the medium and the jet~\cite{Sapeta:2007}. 
\end{enumerate}

We note that even if the {\it average} longitudinal jet multiplicity distribution softens,
it may be possible that high-$p_T$ triggered particle correlations remain insensitive
to the properties of the medium. This is so, for instance, if the high-$p_T$ trigger 
should select the subset of parton fragmentation patterns, that escaped
with a negligible medium modification e.g. due to a surface bias effect. Similar
remarks apply to the transverse jet multiplicity distribution and hadrochemical composition.

\subsection{Jet-like particle correlations and a potential all-or-nothing mechanism}

There is a class of measurements,  in which a 
trigger hadron of high transverse momentum $p_T^{\rm trig}$ is correlated with associated
hadrons as a function of their transverse momentum $p_T^{\rm assoc}$ and
their azimuthal angle $\Delta \phi$ with respect to the trigger particle. We call such correlations ``jet-like'', if
$p_T^{\rm assoc}$ is relatively large, $2\mbox{ GeV}/c <  p_T^{\rm assoc} < p_T^{\rm trig}$, say.
For a first theoretical work on jet-like correlations, see e.g.~\cite{Majumder:2004wh}.
The generic trends seen in such correlation functions at RHIC are:
\begin{enumerate}
\item
\underline{Near-side jet-like particle correlations in Au-Au are independent of centrality and}\\ 
\underline{similar to those in $pp$ or d-Au.}\\
In $pp$ and Au-Au collisions at RHIC, near-side (i.e. small $\Delta \phi$) two-particle 
jet-like correlations show an enhancement characteristic of hard-scattering processes.
Compared to $pp$ collisions, the yield of high-$p_T$ trigger particles decreases by a factor 
$\sim 5$ from peripheral to central Au-Au collisions at RHIC. In contrast, jet-like two-particle
correlations do not show a significant centrality dependence. For sufficiently high
threshold trigger $8\mbox{ GeV}/c< p_T^{\rm trig}< 15\mbox{ GeV}/c$, the yield and 
$\Delta\phi$-width of the 
near-side distribution is insensitive to the centrality of Au-Au collisions, and coincides
with the measurement in d-Au collisions~\cite{Adams:2006yt}. The same has been
observed for lower trigger thresholds~\cite{Adler:2002tq}. Also other features of 
jet-like $p_T$-triggered correlation functions, such as the ratio of like-sign to unlike-sign
pairs in jet-like correlations~\cite{Adler:2002tq}, do not show any centrality dependence
and are consistent with the data found in $pp$ collisions. 
\item
\underline{Back-side jet-like particle correlations decrease in yield with increasing 
centrality,} \\ \underline{but keep approximately the same width.}\\
For intermediate $p_T$ triggers ($4\mbox{ GeV}/c < p_T^{\rm trig} < 6$~GeV/$c$) at RHIC, the associated
particle yield for $p_T^{\rm assoc} > 2$~GeV/$c$ disappears as a function of 
centrality~\cite{Adler:2002tq}. If one raises the trigger threshold to higher values
($8\mbox{ GeV}/c < p_T^{\rm trig}< 15$~GeV/$c$), then the back-side jet-like structure reappears again, 
but the yield strongly decreases with centrality. The back-side structure shows no sign of azimuthal 
broadening~\cite{Magestro:2005vm,Adams:2006yt}.
\end{enumerate}
The above features are qualitatively consistent with a schematic all-or-nothing mechanism, 
based on the following picture: If a hadron is triggered on with a high $p_T^{\rm trig}$, then
it is the leading fragment of a parton shower, which propagated essentially unperturbed 
through the medium (``complete survival of entire jet structure''). On the other hand, if the 
parton shower is significantly perturbed by the medium, then the energy of the leading 
fragment is degraded to such an extent, that it becomes unlikely to find this fragment in a 
high-$p_T$ trigger bin (``no survival at all''). This all-or-nothing picture may
be regarded as the most extreme form of a trigger bias: the trigger selects the subclass of
unmodified parton fragmentation patterns and the medium-modification establishes itself
solely in the reduced yield. In this way, this all-or-nothing picture accounts for the strongly reduced
yield of high-$p_T$ trigger particles, characterized e.g. by the nuclear modification factor, 
as well as for the suppression of the back-side yield. It can also account for the absence of 
broadening in both the near-side and the away-side peaks by arguing that the particle pairs 
entering the jet-like correlation function 
belong to parton showers which escaped the medium essentially without interaction and thus
without signs of medium-induced broadening. The picture is also qualitatively consistent
with finer features seen in the data, such as the observation that for
near-side correlations, the particle yield as a function of the effective fragmentation 
variable $z_T=p_T^{\rm assoc}/p_T^{\rm trig}$ is the same in d-Au and Au-Au, independent
of centrality; on the away-side, the particle yield decreases with centrality but shows the same
$z_T$-slope~\cite{Magestro:2005vm}. 

Can such an all-or-nothing mechanism be consistent with the dynamics of QCD radiation physics?
To address this question, one may note first that for a steeply falling partonic $p_T$-distribution, 
it is conceivable that {\it all\/} high-$p_T$ trigger bins are dominated by hadrons, whose parent 
partons suffered no medium-induced parton energy loss~\cite{Baier:2001yt}. 
In other words: while hadrons, whose 
parents suffered some medium-induced energy loss must end up in some $p_T$-bin, they can 
--- for steeply falling distributions --- always be shifted to an abundantly populated lower $p_T$-bin, 
in which their yield is statistically negligible. Recent implementations of radiative parton energy 
loss can account at least qualitatively for this possibility by two 
features~\cite{Eskola:2004cr,Dainese:2004te,Renk:2006sx}: first, even for dense
systems, recent models allow for a sizeable finite probability that the parton shower propagates 
unperturbed through the medium. Second, the distribution of leading fragments in the parton 
shower turns out to be very fragile, once the parton shower has interacted with the medium.
In this way, current model implementations contain the main ingredients needed for implementing
a strong surface bias, which may underlie the all-or-nothing mechanism sketched above.
In model studies, one has also addressed more refined questions, such as whether the 
surviving yield in the away-side correlation arises predominantly from particle pairs emitted 
tangentially to the surface of the collision region~\cite{Loizides:2006cs,Renk:2006pk}, so that 
neither the trigger nor the associated recoil particle traverses a significant amount of matter. 

The all-or-nothing mechanism outlined here is a working hypothesis, which finds some support
in RHIC data and current model analyses. If true, it is a dramatic illustration that jet-like particle
correlations fall short of characterizing the distributions of quenched jets, simply because they
trigger mainly on the small fraction of unquenched survivor jets. To refine this all-or-nothing
mechanism (or rather: to replace it by a picture which allows for gradual manifestations of parton 
energy loss on jet-like correlations),  one should study in particular correlations with lower
$p_T^{\rm assoc}$. This is so, since the trigger particle of $p_T^{\rm trig}$, to the extent to
which it does not arise from a medium-independent fragmentation, should be accompanied
by an increased associated yield at sufficiently small  $p_T^{\rm assoc}$. At RHIC, lowering
$p_T^{\rm assoc}$ below 2~GeV/$c$ for Au-Au collisions, one has observed indeed an enhanced
associated yield with clear indications of broadening of the away-side peak. However, the
kinematic range $p_T^{\rm assoc} < 2$~GeV/$c$ is difficult to disentangle from the large underlying event
multiplicity and it may be affected by other mechanisms, see section~\ref{sec73} below. 

A trigger-biased class of jet measurements, which shows
medium-modifications of associated jet multiplicity only below $p_T^{\rm assoc} < 2$~GeV/$c$,
provides arguably only limited access to a study of the entire quenched jet fragmentation ---
except, of course, if one could demonstrate that this trigger bias is unimportant and that these
jet-like correlations are characteristic for the {\it average} medium-modified parton shower. 
The wider kinematic reach of heavy ion collisions at the LHC may provide means to this end. 
A jet of $E_{\rm T} = 200$~GeV has on average $\approx 7$ charged hadrons with 
$p_T^{\rm assoc} > 5$~GeV/$c$. Although jet-like correlations based on single trigger particles
will bias significantly the average jet fragmentation pattern, one expects qualitatively that
the distribution of associated particles should show imprints of medium-modifications
(namely signs of $p_T$-broadening and enhanced yield) in a wider range of $p_T^{\rm assoc}$, 
which can be disentangled more clearly from the underlying event multiplicity. However, 
this qualitative expectation is not yet supported by model studies. 

\subsection{The pedestal, the ridge, the Mach cone and all that ...}
\label{sec73}
High-$p_T$ triggers affect the underlying event in hadronic collisions. For 
instance, in comparison to minimum bias data, triggering on a high-$p_T$ hadron 
in a $pp$ collision increases the soft event multiplicity by a factor of order 
$\simeq 2$. The hard parton sits on top of a ``pedestal'', which is wide in rapidity~\cite{Arnison:1983gw}. 
Within perturbation theory, such a pedestal may be expected, since large $Q^2$-processes are 
accompanied by initial state radiation, which is broad in rapidity and which will manifest itself in additional low-$p_T$ hadrons. Since
this initial state radiation moves over long ranges with the beam fragments, non-perturbative 
physics may play an important role as well. Also, multiple parton 
interactions may contribute to the pedestal effect~\cite{Sjostrand:1987su}.

The pedestal observed in high-$p_T$-triggered hadron 
collisions is the prototype of a phenomenon, which is clearly related to the 
presence of a high-$Q^2$ process, but which is not due to final state parton fragmentation. 
As such, the pedestal is a structure in the low-$p_T$ trigger-associated particle yield, which 
one cannot expect to reproduce in a model that superimposes a high-$p_T$ final state 
fragmentation pattern on the multiplicity distribution of a minimum bias event. The state of the 
art of modelling high-$p_T$ phenomena in heavy ion collisions is of the latter type, and one 
wonders whether there are --- like the pedestal in $pp$ --- characteristic features in heavy ion 
collisions, which one misses in models superimposing medium-modified hard processes on
minimum bias soft background. 

One candidate for such a feature is the ``ridge'': a trigger particle is accompanied by 
additional associated hadronic activity in some range of intermediate $p_T^{\rm assoc}$
at the {\it near-side only}. This additional multiplicity is wide in rapidity but, unlike the pedestal, 
it is not balanced by a similar amount of activity in the same range of $p_T^{\rm assoc}$
on the away-side. This phenomenon may arise e.g. in a picture~\cite{Voloshin:2003ud}, 
in which the pedestal is
embedded in a transverse flow field. Namely, triggering on a high-$p_T$ particle, one
selects an interaction point which will preferably lie away from the centre of the collision 
region towards the direction of the trigger $p_T$. At this point in the transverse plane,
collective transverse flow is also expected to point in the direction of the trigger $p_T$.
So, any additional initial state hadronic activity, associated with this trigger, may be
expected to be transported by transverse flow towards the near-side. 

The above is but one, albeit speculative, illustration that  if one aims at studying values of
$p_T^{\rm assoc}$ comparable to those in the bulk multiplicity, the study of medium-modified jet
measurements cannot be limited to the study of medium-modified final state parton 
fragmentation patterns on top of minimum bias events. For low $p_T^{\rm assoc}$, it becomes
difficult to establish which part of the additional hadronic activity emerges from the fragmentation 
and energy loss of a hard final state parton. We note that also the much-discussed,
broad structures in the away-side correlations, which have been suggested to indicate
the appearance of Mach cones, do not persist for higher $p_T^{\rm assoc}$, but are only seen
in a rather narrow range of low transverse momentum. Radial flow, anisotropic flow, initial state
radiation and trigger bias effects may all affect characteristic features of associated particle 
distributions in this low $p_T^{\rm assoc}$-regime. 
Heavy ion collisions at the LHC may
help to clarify the dynamical understanding of such soft structures related to high-$p_T$ trigger 
particles, since the hadronic activity in both the incoming and the outgoing state is 
expected to increase significantly with the trigger $p_T$, and may manifest itself in a wider range
of $p_T^{\rm assoc}$.

\section{Connecting Heavy Ion Phenomenology with QCD}
 
How collective phenomena emerge from the fundamental laws of elementary particle physics
is a multi-faceted question, which in the range of extreme matter densities, where physics
is determined by partonic degrees of freedom, has been addressed in different theoretical 
approaches. Historically, high-temperature QCD equilibrium dynamics, studied non-perturbatively
in lattice calculations or perturbatively in finite temperature field theory, has been the first 
theoretical approach with the potential of connecting heavy ion phenomenology with first
principles of QCD. In particular, the most dramatic collective phenomenon, expected in
finite temperature QCD, namely the phase transition to a quark gluon plasma at a critical
temperature and baryochemical potential, has been firmly established in lattice QCD. By
now, these techniques are applied to many questions of phenomenological relevance at the LHC,
including quarkonium suppression, the medium-modification of spectral functions,
dissipative transport coefficients, and fluctuation 
measurements~\cite{Laermann:2003cv,Karsch:2003jg}.

On the other hand, heavy ion phenomenology has established over the last two decades 
strong indications that effects of directed collective motion are at least as important for
understanding the dynamics of heavy ion collisions, than effects of random thermal motion. These
two concepts, collective dynamics and local equilibrium, can coexist. 
Indeed, the modelling of heavy ion collisions in terms of perfect fluid dynamics illustrates 
the extent to which a mesoscopic system with extreme position-momentum
gradients may still maintain local thermal equilibrium. In the discussion of this hydrodynamic 
modelling, the emphasis has shifted gradually from fundamental tests of QCD thermodynamics 
(namely whether the QCD phase transition and its order leave traces in the dynamical evolution) 
to fundamental tests of QCD hydrodynamics  (namely the test of dissipative properties of the
matter, such as viscosities). There are many reasons for this gradual shift of focus at collider 
energies, starting with the observation that at RHIC and LHC energies one likely overshoots 
the critical energy density significantly, and ending with the notorious problem of identifying 
``unambiguous'' signatures of the QCD phase transition. In principle, the characterization of
hydrodynamic features in heavy ion collisions provides an opportunity of connecting heavy
ion phenomenology to first principles in QCD, since both properties of the QCD phase
transition and dissipative transport coefficients are calculable directly from the QCD Lagrangian.
In practice, however, one prerequisite for exploiting these opportunities is a very good 
experimental and theoretical control over the ``perfect fluid baseline'' on top of which one 
aims at establishing dissipative properties. 
Our discussion in sections~\ref{s:hadrochemistry}, \ref{s:anisotropic_flow} and 
\ref{s:femtoscopy} also identified how measurements at the LHC can help to 
establish whether conditions close to this perfect fluid baseline are realized 
at the LHC {\it and\/} whether they were realized at RHIC.

It would be an unwanted bias to limit the study of ``soft'' physics at the LHC to manifestations 
of QCD thermo- and hydrodynamics. In this review, we deliberately started from the observation
that several apparently generic trends in the existing data (e.g. in multiplicity distributions and
collective flow) have not yet found a satisfactory explanation. Agnostic extrapolations of these 
trends to the LHC are at odds with the extrapolation of current models, be it hydrodynamics or 
saturation physics. This indicates that LHC will be a discovery machine also in the soft physics sector. 
In particular, the support for an interpretation of data in terms of hydrodynamics or saturation 
physics would be strengthened qualitatively, if one discovered at the LHC 
deviations from the so far apparently generic trends, which are characteristic for the currently
advocated dynamical models (such as a mild but distinct power-law $\sqrt{s_{_{NN}}}$ increase 
of event multiplicity, or a deviation of $v_2$ from $\ln\sqrt{s_{_{NN}}}$-scaling). On the other
hand, a confirmation of these trends may prompt us to reassess our understanding of the soft
matter produced in heavy ion collisions at both LHC {\it and} RHIC.

LHC will also be a discovery machine outside the soft physics sector at 
mid-rapidity. 
This is so mainly because of the logarithmically wide range in transverse and 
longitudinal momentum, which opens up at 30 times higher 
$\sqrt{s_{_{NN}}} = 5.5$ TeV. 
As discussed in sections~\ref{s:high_pT} and \ref{s:high_pT2}, the resulting 
abundance of hard processes at the LHC provides many novel tools for probing the
produced soft matter. 
The prerequisite for exploiting this opportunity is a very good experimental and
theoretical control over how the medium modifies hard processes due to 
interactions in the final and initial state. 
Our discussion in section~\ref{s:high_pT2} identified how measurements at the 
LHC can improve this control, in particular by extending jet quenching studies 
significantly beyond the analysis of medium-modified leading fragments. 
This is likely to refine our understanding of hard probes at the LHC {\it and\/}
at RHIC. 

It would be an unwanted bias to limit the study of hard physics at the LHC to its use as ``hard probes''.
With the significantly wider kinematic reach, heavy ion physics gains experimental access to other
fundamental properties of QCD. In particular, medium modifications of the QCD-evolution in both 
$Q^2$ (mainly via the transverse momentum dependence) and $\ln 1/x$ (mainly via the rapidity 
and $\sqrt{s_{_{NN}}}$-dependence)
become testable at the LHC. Again, characteristic deviations of LHC measurements from the agnostic
extrapolations discussed here may provide some of the cleanest possibilities of identifying and
ultimately quantifying the manifestations of medium-dependent QCD evolution. 


\end{document}